%% file: ms.tex
\documentclass[12pt,preprint]{aastex}

\newcommand{\chandra}{{\sl Chandra}}
\newcommand{\eg}{e.g.}
\newcommand{\etal}{et~al.}
\newcommand{\ie}{i.e.}
\newcommand{\ltapprox}{\hbox{\raise0.5ex \hbox{$<$}
  \kern-1.1em \lower0.5ex \hbox{$\sim$}}}
\newcommand{\snr}{SN~1006}

\slugcomment{ApJ, accepted}

\shorttitle{A Curved Synchrotron Spectrum in \snr}
\shortauthors{}

\begin{document}

\title{Evidence of a Curved Synchrotron Spectrum in the Supernova Remnant
\snr}

\author{G.\ E.\ Allen and J.\ C.\ Houck}
\affil{MIT Kavli Institute for Astrophysics and Space Research, Cambridge,
MA 02139; gea@space.mit.edu; houck@space.mit.edu} 

\and

\author{S.\ J.\ Sturner}
\affil{Astroparticle Physics Laboratory, Code 661, NASA Goddard Space Flight
Center, Greenbelt, MD 20771; sturner@swati.gsfc.nasa.gov}



\begin{abstract}

A joint spectral analysis of some \chandra\ ACIS X-ray data and Molonglo
Observatory Synthesis Telescope radio data was performed for 13 small
regions along the bright northeastern rim of the supernova remnant \snr.
These data were fitted with a synchrotron radiation model.  The nonthermal
electron spectrum used to compute the photon emission spectra is the
traditional exponentially cut off power law, with one notable difference:
The power-law index is not a constant.  It is a linear function of the
logarithm of the momentum.  This functional form enables us to show, for the
first time, that the synchrotron spectrum of \snr\ seems to flatten with
increasing energy.  The effective power-law index of the electron spectrum
is 2.2 at 1~GeV (\ie, radio synchrotron--emitting momenta) and 2.0 at about
10~TeV (\ie, X-ray synchrotron--emitting momenta). This amount of change in
the index is qualitatively consistent with theoretical models of the amount
of curvature in the proton spectrum of the remnant.  The evidence of
spectral curvature implies that cosmic rays are dynamically important
instead of being ``test'' particles. The spectral analysis also provides a
means of determining the critical frequency of the synchrotron spectrum
associated with the highest-energy electrons.  The critical frequency seems
to vary along the northeastern rim, with a maximum value of
$1.1^{+1.0}_{-0.5} \times 10^{17}$~Hz.  This value implies that the electron
diffusion coefficient can be no larger than a factor of $\sim$4.5--21 times
the Bohm diffusion coefficient if the velocity of the forward shock is in
the range 2300--5000~km~s$^{-1}$.  Since the coefficient is close to the
Bohm limit, electrons are accelerated nearly as fast as possible in the
regions where the critical frequency is about $10^{17}$~Hz.

\end{abstract}

\keywords{
   acceleration of particles ---
   cosmic rays ---
   ISM: individual (\snr) ---
   radiation mechanisms: nonthermal ---
   supernova remnants ---
   X-rays: general
   }



\section{Introduction}
\label{int}

Several observational and theoretical clues support the suggestion that
Galactic cosmic rays, up to an energy of 100~TeV \citep{lag83} or more
\citep{jok87,vol88,bel01}, are accelerated predominantly in the shocks of
supernova remnants.  If Galactic supernovae occur at an average rate of one
event every 30 years, then an average supernova remnant must transfer about
10\% \citep{dru89} of the initial $10^{51}$~ergs of kinetic energy of the
ejecta to cosmic rays. In this case, the cosmic-ray energy density may be
large enough to affect the structure of the shock (\ie, cosmic rays may not
be mere ``test'' particles).

Three consequences of a large cosmic-ray pressure are potentially
observable.  One consequence is that the ambient material is slowed before
it crosses the subshock.  This effect reduces the temperature of the shocked
gas \citep{che83,ell00,dec00}. An upper limit on the temperature is provided
by the well-known relation between the shock speed $v_{s}$ and the postshock
temperature of a gas whose ratio of specific heats $\gamma = \frac{5}{3}$:
\begin{equation}
kT_{i}
  \le
  \frac{3}{16} m_{i} v_{s}^2,
  \label{eqn1}
\end{equation}
where $k$ is Boltzmann's constant and $m_{i}$ and $T_{i}$ are the mass and
immediate postshock temperature, respectively, of particle species $i$.
Since kinetic energy is transferred to cosmic rays at the expense of the
thermalization of the shocked gas, the equality is appropriate only in the
limit that the cosmic-ray energy density is negligible.  \cite{hug00}
analyzed the transverse motion of the X-ray--emitting material (mostly
reverse-shocked oxygen and neon) in the supernova remnant 1E~0102.2$-$7219
and inferred a forward-shock velocity $v_{s} =
6200^{+1500}_{-1600}$~km~s$^{-1}$.  In this case, the temperature of the
shocked protons must be lower than the right-hand side of
equation~(\ref{eqn1}) or else the fitted electron temperature will be too
low to be explained by Coulomb heating \citep{hug00}.  However, the proton
temperature is sensitive to the shock velocity ($T \propto v_{s}^{2}$), and
a variety of speeds are reported for 1E~0102.2$-$7219. \cite{fla04} report
that the \chandra\ High Energy Transmission Grating \ion{Ne}{10} line
emission data are consistent with a model that includes radial \ion{Ne}{10}
velocities up to $1800 \pm 450$~km~s$^{-1}$. \citet{fin06} analyzed several
features in \ion{O}{3} images.  The mean transverse velocity of the features
is $2000 \pm 200$~km~s$^{-1}$. \citet{eri01} report that the motion of some
\ion{O}{3}--emitting material is best described by a radial \ion{O}{3}
velocity of about 1800~km~s$^{-1}$. Since these velocity results differ and
since none of the measurements provides a direct measure of the velocity of
the forward shock, the claim that a significant fraction of the internal
energy in 1E~0102.2$-$7219 has been transferred to cosmic rays requires
additional support.

A second consequence is that the total compression ratio is larger than 4
\citep{ell91,ber99}. Since cosmic rays slow the upstream material before it
crosses the subshock, the velocity of the shocked material relative to the
subshock is reduced.  As a result, the contact discontinuity is closer to
the subshock than it would be in the absence of a large cosmic-ray pressure.
\citet{war05} report that in Tycho's supernova remnant the mean ratio of the
radius of the contact discontinuity to the radius of the forward shock is
$0.93 \pm 0.02$. This separation corresponds to a total compression ratio $r
= 5.1_{-1.0}^{+1.9}$.  Similarly, \citet{cas08a} report that for the
southeastern rim of \snr\ the ratio varies from a value of
$0.91^{+0.03}_{-0.02}$ (\ie, $r = 4.0^{+2.2}_{-0.5}$) between the bright
X-ray synchrotron--emitting filaments to a value of 1 (\ie, $r = \infty$)
along the filaments. The mean value between the filaments is $0.96 \pm 0.03$
(\ie, $r = 9.0^{+25}_{-3.6}$). These authors explore ways in which the
location of the contact discontinuity can approach the location of the
forward shock, but they cannot explain why these two features curiously
appear to be coincident with one another along the filaments.
\citet{kse05a}, who also studied \snr, infer a mean compression ratio $r =
5.2$.

A third consequence of a large cosmic-ray pressure is that the shock
transition region is broadened or ``smeared out'' \citep{ell91,ber99}.  Only
the subshock has a short transition length.  In this case, low-energy cosmic
rays, which have relatively small diffusion lengths, experience only a
portion of the velocity gradient as they scatter back and forth across the
subshock. Higher energy particles, which have relatively large diffusion
lengths, experience a larger portion (or all) of the velocity jump. Since
the rate of energy gain increases as the velocity difference increases,
higher energy particles gain energy faster than lower energy particles. As a
result, cosmic-ray spectra do not have power-law distributions.  The spectra
flatten with increasing energy \citep{bel87,ell91,ber99}. \cite{jon03} and
\cite{vin06} report evidence of curvature in the synchrotron spectra of
Cas~A and RCW~86, respectively.  One possible explanation for such curvature
is that the cosmic-ray electrons producing the synchrotron emission have
curved spectra.

Here we describe the results of an analysis of some radio and X-ray
synchrotron data, which suggest that the synchrotron spectrum of \snr\ might
be curved.  The data, assumptions, and analysis techniques are described in
\S\ \ref{dat}. The results of the analysis are discussed in \S\ \ref{dis}.
Our conclusions are summarized in \S\ \ref{con}.



\section{Data and analysis}
\label{dat}

X-ray spectra of the filaments along the northeastern rim of \snr\ were
obtained by analyzing 68~ks of data from the 2000 July 10--11 {\sl
Chandra}\footnote{For more information about the {\sl Chandra X-ray
Observatory} refer to the {\sl Chandra} Proposers' Observatory Guide, which
is available at http://asc.harvard.edu/proposer/POG/index.html.} observation
of the source. The results of previous analyses of the data from this
observation were published by \citet{lon03} and \citet{bam03}. During the
observation, the telescope was pointed at a location near the middle of the
northeastern rim ($\alpha = 15^{\rm h} 03^{\rm m} 51.56^{\rm s}$, $\delta =
-41\arcdeg 51\arcmin 18.8\arcsec$; J2000). The X-rays were detected using
the Advanced CCD Imaging Spectrometer (ACIS), which includes a $2 \times 2$
(ACIS-I) and a $1 \times 6$ (ACIS-S) array of CCDs.  Six of these 10
detectors (ACIS-I2, -I3, -S1, -S2, -S3, and -S4) were used for the
observation. Each $1024\ {\rm pixel} \times 1024\ {\rm pixel}$ CCD has a
field of view of $8.4\arcmin \times 8.4\arcmin$. The angular resolution of
the {\sl Chandra} mirrors and ACIS varies over the observed portion of \snr\
from about $0.5\arcsec$ at the aim point to $20\arcsec$ for a region that is
$17.5\arcmin$ off-axis.  The on-axis effective area for the mirrors and
ACIS-S3 has a maximum of about 720~cm$^{2}$ at 1.5~keV and is greater than
10\% of this value for energies between about 0.3 and 7.3~keV. The
fractional energy resolution (FWHM/$E$) between these energies ranges from
about 0.4 to 0.03, respectively.  The sensitive energy bands and energy
resolutions of the other five CCDs used for \snr\ are typically narrower and
worse, respectively, than the energy band and resolution of ACIS-S3.

The ACIS data were filtered to remove the events that (1) have ${\rm GRADE}
= 1$, 5, or 7, (2) have one or more of the STATUS bits set to 1 (except for
events that have only one or more of the four cosmic-ray ``afterglow'' bits
set), (3) occur on a bad pixel or column, (4) are part of a serial readout
streak on ACIS-S4, or (5) occur in the time interval during which the mean
background count rate was more than twice the nominal rate (\ie, frames
20515--21294).  An image of the 66~ks of ``good'' data is displayed in
Figure~\ref{fig1}.  The 13, 100~pixel $\times$ 100~pixel ($49 \arcsec \times
49 \arcsec$) boxes are the regions of the bright, northeastern filaments
used for the X-ray spectral analysis. The PHA spectra, ARFs, and RMFs for
each region were created using version 2.3 of the CIAO\footnote{For more
information about the CIAO analysis tools, refer to
http://asc.harvard.edu/ciao.} tools {\tt dmextract}, {\tt mkarf}, and {\tt
mkrmf}, respectively.

The X-ray analysis was performed using data in the energy band 2--7~keV. The
higher energy data were excluded because they are dominated by background
events.  The lower energy data were excluded for three reasons. First, a
small, but noticeable, amount of thermal emission is evident at the lower
energies.  Therefore, inclusion of the lower energy data requires a spectral
model that includes a thermal emission component.  The extra parameters
required to fit the thermal emission make the spectral fits unnecessarily
complicated.  By ignoring the data below 2~keV, nearly all of the thermal
emission is excluded \citep{bam03}.  As displayed in Figure~\ref{fig1} (and
as originally reported by Koyama et~al.\ 1995), the high-energy nonthermal
emission is concentrated along the rim. The second reason is that the ACIS
detectors are not well calibrated at energies below about 0.7~keV.  The last
reason is that the absorption column density can be frozen at some preferred
value to simplify the fitting process.  Our preferred value is $n_{\rm H} =
6.0 \times 10^{20}$~cm$^{-2}$, which is consistent with the results of
\citet{sch96} and \cite{all01}.

The radio data used for the spectral analysis include the compilation of
flux density measurements listed in Table~\ref{tab1}.  These measurements
are based on analyses of the emission from the entire supernova remnant. In
the cases where no flux density uncertainties are reported, the
uncertainties are assumed to be 10\% of the reported flux densities.  The
results of the spectral analysis are insensitive to the actual fraction used
at least for fractions in the range 5\%--20\%.  The values listed in
Table~\ref{tab1} are plotted in Figure~\ref{fig2}.  The dashed line in this
figure is the result obtained when only the radio data are fitted with a
power-law model: $S(\nu) = 17.6^{+6.3}_{-4.8}[\nu/(1~{\rm
GHz})]^{-\alpha}$~Jy, where $\alpha = 0.60^{+0.08}_{-0.09}$. Note that the
uncertainties for the power-law index and normalization, which are quoted at
the 90\% confidence level, are not independent.  This result is an update of
and consistent with the result reported by \citet{all01}: $S(\nu) = 17.9 \pm
1.1 [\nu / (1\ {\rm GHz})]^{-\alpha}$~Jy, where $\alpha = 0.57 \pm 0.06$.
The dotted line in Figure~\ref{fig2} is a power-law model with the
parameters reported by \citet{gre06}: $S(\nu) = 19[\nu/(1~{\rm
GHz})]^{-0.6}$~Jy.

The radio spectrum for the entire remnant is used to construct radio spectra
for each region because there is little evidence of radio spectral
variability.  For example, a comparison of the images of \citet{ste77} and
\citet{gar65} suggests that the slopes of the 0.408--2.7~GHz radio spectra
of the northeastern and southwestern rims are about the same. It is possible
that the slope may vary along each rim, in which case the best-fit amounts
of curvature for individual regions may be inaccurate, but the mean amount
of curvature for the 13 regions is relatively insensitive to such
variations. Two sets of radio data were used with each X-ray spectrum. Both
sets are obtained by multiplying the remnant-integrated flux densities and
uncertainties of Table~\ref{tab1} by scaling factors $\zeta$.  One set of
factors (Tables~\ref{tab2} and \ref{tab3}) is obtained by dividing the
843~MHz flux densities for the 13 regions marked with black squares in
Figure~\ref{fig3}, from the Molonglo Observatory Synthesis Telescope (MOST),
by the 843~MHz flux density for the entire remnant.  These regions are the
same ones used to obtain the X-ray spectra (Fig.~\ref{fig1}).  The second
set of factors (Tables~\ref{tab4} and \ref{tab5}) is obtained in the same
way, except that the flux densities for the 13 red squares are used instead
of the flux densities for the black squares. The red squares are the regions
where the flux densities peak.  That is, the flux density along a line from
the center of the remnant through the center of a black square peaks at the
location of the center of the corresponding red square. Although both sets
of flux densities represent local averages because the half-power beamwidth
associated with the 843~MHz MOST image ($44'' \times 66''$; see
Fig.~\ref{fig3}) is comparable to the size of the regions ($49'' \times
49''$), they are reasonable estimates of the minimum ({\sl black}) and
maximum ({\sl red}) flux densities for the 13 regions.  Therefore, using
these two sets of radio spectra provides a means of obtaining upper
(Table~\ref{tab2}) and lower (Table~\ref{tab4}) limits on the amount of
curvature for each region.

Version 1.4.8 of the spectral fitting package ISIS\footnote{For more
information about ISIS, see http://space.mit.edu/cxc/isis.} \citep{hou00}
was used to fit the data for each region with a model that includes a
synchrotron radiation component for the X-ray and radio data and an
interstellar absorption component for the X-ray data (the XSPEC model {\tt
wabs}, with the relative abundances of Anders \& Grevesse 1989). Since the
2--7~keV emission is dominated by nonthermal emission, no thermal X-ray
component is included.

The synchrotron spectra are based on a nonthermal electron spectrum of the
form
\begin{equation}
\frac{dn}{dp}
  =
  A \left( \frac{p}{p_{0}} \right)^{-\Gamma + a \log
    \left( \frac{p}{p_{0}} \right)}
  {\rm exp} \left( \frac{p_{0} - p}{p_{m}} \right),
  \label{eqn2}
\end{equation}
where $n$ is the electron number density, $p = \gamma m v$ ($\gamma$, $m$,
and $v$ are the Lorentz factor, rest mass, and velocity of a particle,
respectively), $A$ is the number density at $p = p_{0}$ (in units of
$p_{0}^{-1}$ cm$^{-3}$), $p_{0} = 1$~GeV~$c^{-1}$ ($c$ is the speed of
light), $\Gamma$ is the differential spectral index at $p = p_{0}$, $a$ is
the spectral ``curvature,'' the logarithm is base 10, and $p_{m}$ is the
exponential cutoff (or ``maximum'') momentum.  This form is the same as the
standard power law with an exponential cutoff, except for the term $a \log
\left( p / p_{0} \right)$.  Introduction of this term produces an effective
spectral index, $\Gamma_{\rm eff} = \Gamma - a \log \left( p / p_{0}
\right)$, that is a linear function of the logarithm of the momentum.  For
example, at momenta $p$ of $10^{0} p_{0}$, $10^{1} p_{0}$, $10^{2} p_{0}$,
and $10^{3} p_{0}$, the effective differential spectral indices are
$\Gamma$, $\Gamma - a$, $\Gamma - 2a$, and $\Gamma - 3a$, respectively. If
$a > 0$, then the spectrum flattens with increasing momentum.  If $a < 0$,
then the spectrum steepens with increasing momentum. If $a = 0$, then the
spectrum has no curvature and equation~(\ref{eqn2}) reduces to the standard
power law with an exponential cutoff.  An exponential cutoff is used in
equation~(\ref{eqn2}) for simplicity. Note that \citet{ell01},
\citet{uch03}, \citet{laz04}, and \cite{zir07a} use a more general form for
the cutoff (exp$[- \left( p / p_{m} \right)^{s} ]$, where $s = \case{1}{4}$,
$\case{1}{2}$, 1, or 2) and that \citet{pro04} performed a detailed study of
the shape of the cutoff for particles experiencing diffusive shock
acceleration.

The synchrotron spectrum is computed by evaluating the following integral
expression to obtain the differential photon flux\footnote{For more
information about this model, see \citet{hou06}.} (in SI units):
\begin{equation}
\frac{dF}{d(h\nu)}
  =
  \frac{V_{S}}{4 \pi d^{2}}
  \frac{\sqrt{3} e^{3} B}{2 \epsilon_{0} m c h^{2} \nu}
  \int dp \frac{dn}{dp}
  \int_{0}^{\pi} d\theta \sin^{2}\!\theta
  \frac{\nu}{\nu_{c}}
  \int_{\nu / \nu_{c}}^{\infty} dx K_{5/3}
    \left( x \right),
  \label{eqn02}
\end{equation}
where $V_{S}$ is the volume of the synchrotron-emitting region, $d$ is the
distance of the source from Earth, $e$ is the unit of electric charge, $B$
is the total magnetic field strength (not $B_{\perp}$), $\epsilon_{0}$ is
the permittivity of free space, $h$ is Planck's constant, $\nu$ is the
frequency of an emitted photon, $\theta$ is the pitch angle between the
electron momentum and magnetic field vectors, $K_{5/3}$ is an irregular
modified Bessel function of the second kind, and
\begin{equation}
\nu_{c}
  =
  \frac{3 e}{4 \pi m^{3} c^{4}}
  \left( p^{2}c^{2} + m^{2}c^{4} \right)
  B \sin \theta
  \label{eqn03}
\end{equation}
is the critical frequency of the synchrotron spectrum produced by an
electron with a momentum $p$. The frequency at which a synchrotron power
spectrum peaks is $\nu_{p} = 0.286 \nu_{c}$. Hereafter, only the critical
frequency $\nu_{c}$ is used.

We refer to the critical frequency associated with electrons that have a
momentum equal to $p_{m}$ as the ``cutoff critical frequency'' or just the
cutoff frequency to emphasize that it is associated with electrons at the
cutoff momentum of the electron spectrum [$\nu_{m} \equiv \nu_{c} (p =
p_{m})$].  The distribution of pitch angles $\theta$ is expected to be
nearly isotropic because the forward-shock speed (2300--5000~km~s$^{-1}$;
Laming \etal\ 1996; Dwarkadas \& Chevalier 1998; Ghavamian \etal\ 2002) is
much less than the speed of light.  In this case, the mean value of $\sin
\theta$ in equation~(\ref{eqn03}) is about $\pi / 4$,
\begin{equation}
\nu_{m}
  =
  1.26 \times 10^{17}
  \left( \frac{p_{m}}{10\ {\rm TeV}\ c^{-1}} \right)^{2}
  \left( \frac{B}{100\ \mu{\rm G}} \right)\ {\rm Hz},
  \label{eqn04}
\end{equation}
and
\begin{equation}
h \nu_{m}
  =
  0.522
  \left( \frac{p_{m}}{10\ {\rm TeV}\ c^{-1}} \right)^{2}
  \left( \frac{B}{100\ \mu{\rm G}} \right)\ {\rm keV}.
  \label{eqn05}
\end{equation}

During the spectral fitting process, the right-hand side of
equation~(\ref{eqn02}) is computed each time the spectral fitting program
changes the parameters of the model.  The complete set of spectral
parameters includes the spectral index $\Gamma$, the curvature parameter
$a$, the cutoff momentum $p_{m}$, the total magnetic field strength $B$, the
normalization constant $N_{S} = A V_{S} / 4 \pi d^{2}$, and the absorption
column density $n_{\rm H}$. The spectral index is determined by the radio
and X-ray data. The curvature is sensitive to the relative radio and X-ray
fluxes.  Since only two of the three parameters $p_{m}$, $B$, and $N_{S}$
are independent, $p_{m}$ was fixed and the cutoff frequency $\left( \nu_{m}
\propto p_{m}^{2} B;\ {\rm eq.~[\ref{eqn04}]} \right)$ is reported here
instead of the magnetic field strength. This frequency is sensitive to the
shape of the X-ray spectrum. The same normalization is used for both the
X-ray and radio data for a given region because the nonthermal X-ray and
radio emission from the region is assumed to be produced by a common
population of electrons.  The normalization $N_{S}$ is determined by the
radio data.  The absorption column density $n_{\rm H}$ was fixed at the
value $6.0 \times 10^{20}$~cm$^{-2}$ \citep{sch96,all01}.

Four spectral fits were performed for each region. The first pair are
identical to one another except that the curvature parameter is a free
parameter for one (Table~\ref{tab2}) and is fixed at a value of zero (\ie,
no spectral curvature) for the other (Table~\ref{tab3}).  In both cases, the
radio spectra are normalized to the flux densities in the regions that are
cospatial with the X-ray regions (\ie, the regions marked with black squares
in Fig.~\ref{fig3}).  Since the cospatial radio spectra represent lower
limits on the radio fluxes, the amounts of curvature listed in
Table~\ref{tab2} are upper limits. The second pair of fits
(Tables~\ref{tab4} and \ref{tab5}) is identical to the first pair except
that the radio spectra are normalized to the flux densities in the ``peak''
regions (\ie, the regions marked with red squares in Fig.~\ref{fig3})
instead of the cospatial regions. Since the peak radio spectra represent
upper limits on the radio fluxes, the amounts of curvature listed in
Table~\ref{tab4} are lower limits. For each of the four sets of fits, the
best-fit values of the spectral index vary little from region to region,
because the same radio spectral shape was used for each region.  The
best-fit model for region~6, the region with the largest number of counts,
is plotted in Figures~\ref{fig2} and \ref{fig4}.



\section{Discussion}
\label{dis}

It is not immediately obvious whether the cospatial or peak radio fluxes
yield results that are more accurate, because the X-ray and radio
synchrotron morphologies are not the same.  For example, along $49''$-wide
strips that pass through the 13 X-ray regions toward the center of the
remnant, one finds that the X-ray emission peaks at the locations of the
black boxes in Figure~\ref{fig3} and that the radio emission peaks at the
locations of the red boxes. There are at least three possible explanations
for this difference. One explanation is that the TeV electrons that produce
the X-ray synchrotron emission suffer significant synchrotron losses as they
diffuse downstream from the shock. These losses deplete the highest-energy
end of the electron spectrum, causing the X-ray synchrotron emission to
decline in the downstream region. As a result, the X-ray peak is relatively
narrow and confined to a region near the shock. Since synchrotron losses are
negligible for the GeV electrons that produce the radio emission, the radio
emission is not attenuated in the same manner. The radio peaks can be
broader and farther from the shock than the X-ray peaks. Nevertheless, there
is no reason, on these grounds, to use different radio and X-ray extraction
regions, because the radio and X-ray synchrotron emission for a region is
produced by the same population of electrons.  The radio emission is
produced by the GeV portion of the electron spectrum and the X-ray emission
is produced by the TeV segment. Synchrotron losses do not introduce a
spatial separation between the GeV and TeV electrons.

Another explanation for the different X-ray and radio morphologies does
involve an apparent (not actual) spatial separation. The X-ray and radio
instruments do not yield images that have the same spatial resolution. The
size of the spatially dependent point-spread function of the {\sl Chandra}
X-ray telescope is much smaller than the size of the $49\arcsec \times
49\arcsec$ boxes and, hence, is neglected. However, the size of the
beamwidth for the MOST image ($44\arcsec \times 66\arcsec$) is not
negligible, because it is comparable to the size of the boxes. The
implications of the difference in the qualities of the X-ray and radio
images are illustrated in Figure~\ref{fig5}. The black histogram in the top
panel shows the radial X-ray profile for region~6. The red curve in this
panel is obtained by using a Gaussian function to smooth the histogram to
the beamwidth of the radio instrument.  Note that the peak of the red curve
is about $18\arcsec$ downstream from the peak of the histogram, which
suggests that the radio extraction region should be shifted toward the radio
peak by the same amount. Similarly small shifts are obtained for the other
12 regions and for a simple model of a uniformly emitting spherical shell
(Fig.~\ref{fig5}, {\sl bottom}).

A third explanation is that the bright radio-emitting region has expanded in
the 17 intervening years between the time when the MOST data were obtained
and the time of the \chandra\ observation.  Using the mean radio expansion
result of $0.049{\rm \%} \pm 0.014$\%~yr$^{-1}$ \citep{mof93a} suggests that
the radio peak has shifted about $7.6'' \pm 2.2''$.  To compensate for this
expansion, the radio extraction region should be shifted toward the radio
peak by the same amount. Since this shift and the apparent shift described
above are small compared with the size of the extraction region, the
following discussion focuses primarily on the results obtained using the
cospatial radio fluxes.

\subsection{Curvature}
\label{curv}

While the values of $\chi^{2}$ per degree of freedom (dof) are acceptable
for each of the four sets of fits, a comparison of Tables~\ref{tab2} and
\ref{tab3} (and of Tables~\ref{tab4} and \ref{tab5}) reveals that the values
are uniformly lower when curvature is included as a free parameter. For
example, Figure~\ref{fig6} shows the 1, 2, and 3~$\sigma$ confidence
contours for region~6 in the parameter space defined by the electron
spectral index $\Gamma$ and the spectral curvature $a$.  If the cospatial
radio flux density is used, the results for this region suggest that an
uncurved spectrum (\ie, the dashed line at $a = 0 $) can be excluded at
about the 2.7~$\sigma$ confidence level.  Similar results, with varying
degrees of statistical significance, are obtained for the other 12 regions.
Table~\ref{tab2} lists the probability $P_{\Delta \chi^{2}}$ that the
reduction in the value of $\chi^{2}$ when curvature is included as a free
parameter is due to chance. The majority of the probabilities are less than
1\%. Some are much less. The probabilities obtained using the F-test
($P_{F}$) are within a factor of 2 of the values listed in Table~\ref{tab2},
except for regions 1, 2, 3, 5, 6, and 7, where $P_{F} = 0.058$, 0.0035, $9.9
\times 10^{-6}$, 0.0013, 0.0030, and 0.048, respectively. The relatively
large discrepancies between $P_{\Delta \chi^{2}}$ and $P_{F}$ for these
regions arise because the F-test depends on the value of $\chi^{2}/{\rm
dof}$ in addition to the value of $\Delta \chi^{2}$. Note that the values of
$\chi^{2}/{\rm dof}$ are significantly less than 1 for regions~1, 2, and 3
and that regions~5, 6, and 7 have the largest values of $\chi^{2}/{\rm
dof}$. Since the values of $P_{\Delta \chi^{2}}$ are typically larger (\ie,
less significant) than the values of $P_{F}$, only the values of $P_{\Delta
\chi^{2}}$ are listed in Table~\ref{tab2}. Collectively, these probabilities
show that the spectra are significantly better fitted by a curved
synchrotron model than an uncurved one.

The strength of this evidence depends on the location used to obtain the
radio flux density. The radio flux density for a region increases from the
cospatial location to the peak location. Since the appropriate location to
use lies between these two, the cospatial and peak flux densities represent
the lower and upper limits, respectively, on the radio flux density for the
region. The amount of curvature depends on the relative X-ray and radio
synchrotron fluxes.  Therefore, the relatively small cospatial flux density
yields an upper limit on the amount of curvature and the peak flux density
yields a lower limit.  Although the curvature values in Table~\ref{tab4} are
lower limits, even these results favor a curved synchrotron model.  Yet, as
noted above, the upper limit values in Table~\ref{tab2} are expected to be
closer to the actual amount of curvature because the appropriate location to
use for the radio flux density is much closer to the cospatial location than
the peak location.

The evidence of curvature cannot be attributed to an instrumental effect. It
must be associated with the synchrotron spectrum of the source.  For
example, the need for curvature could be eliminated if the effective area of
the X-ray detector system were too low, but the area would have to be in error
by an implausibly large factor of about 4.6. The use of a curved source
model is also compelling in the sense that the values of the best-fit
indices and curvature parameters are consistent with the expected values.
One expectation is that the best-fit index should be consistent with the
radio spectral index.  That is, the best-fit model should fit the radio
spectrum instead of simply intersecting it. The mean value of the index
$\Gamma = 2.221 \pm 0.013$ if curvature is included as a free parameter
(Table~\ref{tab2}). This value is consistent with the index obtained if only
the radio data are fitted with a power-law model ($\Gamma = 2 \alpha + 1 =
2.20^{+ 0.16}_{- 0.18}$; Fig.~\ref{fig2}).  Yet this result is not
surprising, because the fitted value of the index is determined largely by
the shape of the radio spectrum. For comparison, the mean value of the index
$\Gamma = 2.031 \pm 0.017$ if an uncurved model is used (Table~\ref{tab3}).
This value differs significantly from 2.22 because the uncurved index,
unlike the curved index, is sensitive to the relative radio and X-ray
fluxes. While this result seems to be closer to the case of having the model
simply intersect the radio spectrum, an index of 2.03 is still marginally
consistent with the radio data because the radio data have relatively large
uncertainties. Otherwise, the radio data might provide a means of
discriminating between a curved and an uncurved model.

Another expectation is that the value of the curvature parameter should be
positive.  That is, the spectrum should flatten with increasing energy
\citep{bel87,ell91,ber99}.  The results in Table~\ref{tab2} show that the
best-fit values are greater than zero for all 13 regions.  If the spectra
are not curved, then the use of a curvature parameter may not be physically
meaningful and one might expect roughly half of the values to be negative.
If either sign is equally likely, then the chance probability that all 13
values have the same sign is 0.00024 $(2^{-12})$.

A third expectation is that the highest-energy electrons, the ones that
experience the full strength of the shock and that produce the X-ray
synchrotron emission, should have an effective spectral index of 2
\citep{ber02}. If equation~(\ref{eqn2}) is an accurate representation of the
shape of the electron spectrum, then the effective spectral index in the
X-ray band $\Gamma_{X} = \Gamma - a \log \left (p_{X} / p_{0} \right)$,
where $p_{X}$ is the flux-weighted mean momentum of the electrons that
produce the X-ray synchrotron radiation. To compute $p_{X}$ from the fitted
value of $\nu_{m}$ requires knowledge of the magnetic field strength
(eq.~[\ref{eqn04}]).  Since the results of our analysis do not yield an
estimate of the field strength, we assume that $p_{X} = 10$~${\rm TeV}\
c^{-1}$. (The results are rather insensitive to this assumption as long as
$p_{X}$ is within an order of magnitude of 10~${\rm TeV}\ c^{-1}$.) In this
case, $\Gamma_{X} = 2.005 \pm 0.027$. This index and the index obtained
using the uncurved model ($\Gamma = 2.031^{+0.017}_{-0.016}$) are both
consistent with the idea that the highest-energy electrons in \snr\ have a
spectral index of 2.0.

If the effective spectral index of the X-ray synchrotron--producing
electrons $\Gamma_{X} = 2$ (aside from the exponential cutoff), then the
expression for the effective spectral index can be inverted to obtain an
estimate of the expected amount of curvature as a linear function of the
spectral index $\Gamma_{r}$:
\begin{equation}
a_{\rm exp}
  =
  \frac{\Gamma_{r} - 2}{\log \left( p_{X} / p_{r} \right)}
  \label{eqn07}
\end{equation}
where $\Gamma_{r}$ is the effective spectral index in the radio band and
$p_{r}$ is the flux-weighted mean momentum of the electrons that produce the
radio synchrotron emission.  Although the value of $p_{X} / p_{r}$ is
unknown, $p_{X} / p_{r} = \left( \nu_{X} / \nu_{r} \right)^{1/2}$, where
$\nu_{X}$ and $\nu_{r}$ are the critical frequencies associated with $p_{X}$
and $p_{r}$, respectively. Since the X-ray data used here span the range
from 2 to 7~keV, $\nu_{X} \approx 9.0 \times 10^{17}$~Hz, the logarithmic
mid-point of this band. Similarly, $\nu_{r} \approx 660$~MHz, the
logarithmic mid-point of the range from 86~MHz to 5~GHz (Table~\ref{tab1}).
These estimates yield $\log \left( p_{X} / p_{r} \right) = 4.57$. This value
is rather insensitive to the uncertainties in $\nu_{X}$ and $\nu_{r}$. For
example, a change of a factor of 2 in the ratio $\nu_{X} / \nu_{r}$
corresponds to a change of only 3.3\% in $\log \left( p_{X} / p_{r}
\right)$. As shown in Figure~\ref{fig6} ({\sl dotted line}), a relation of
the form $a_{\rm exp} = \left( \Gamma_{r} - 2 \right) / 4.57$ is consistent
with the confidence contours for region~6.  This result is remarkable
because the contours in Figure~\ref{fig6} are sensitive to the relative
X-ray and radio fluxes, while equation~(\ref{eqn07}) has no such dependence.
Had the relative normalizations been significantly different, it is unlikely
that the contours in Figure~\ref{fig6} would have been consistent with the
dotted line. If $\Gamma_{r}$ is accurately represented by the best-fit index
(\ie, if $p_{r} \approx 1$~${\rm GeV}\ c^{-1}$) and if the cospatial radio
fluxes are used, then $a_{\rm exp} = (2.221 - 2) / 4.57 = 0.048$. This
expected value is consistent with the fitted value ($0.054 \pm 0.006$) at
the 90\% confidence level.  If the peak radio fluxes are used, then $a_{\rm
exp} = (2.198 - 2) / 4.57 = 0.043$, which is less consistent with the fitted
value ($0.033^{+0.007}_{-0.008}$). Like the expected radio flux, the
expected amount of curvature lies between the cospatial and peak values. If
the uncertainty associated with the radio flux leads to an uncertainty in
the amount of curvature comparable to the statistical uncertainty, then the
amount of curvature is most likely $0.05 \pm 0.01$.

Not only is the best-fit amount of curvature consistent with
equation~(\ref{eqn07}), it is also consistent with the amount of curvature
predicted for the proton (not electron) spectrum of \snr. For example,
Figure~\ref{fig7} shows the best-fit electron spectra for region~6. The
spectra are not curved below $p = 1\ {\rm GeV}\ c^{-1}$, which was true
during the fitting process, because the available data poorly constrain the
shape of the electron spectrum in this momentum range. The top pair of
dotted and dot-dashed curves are predictions described by \cite{ell00} for
the proton (not electron) spectrum of \snr. These spectra have been
normalized to match the solid black curve at $E = 0.9$~GeV. Since
\cite{ell00} computed their spectrum assuming a spectral index $\Gamma =
2.0$, it is not appropriate to compare the solid curve with the upper pair
of dotted and dot-dashed curves. However, the lower pair of dotted and
dot-dashed curves have been multiplied by the energy-dependent factor $[E /
(0.9\ {\rm GeV})]^{-0.2}$ to match the best-fit spectral index $\Gamma =
2.2$.  Aside from a possible difference in the assumed value of the cutoff
momentum, the lower pair of theoretical curves match the solid black curve
remarkably well.

The evidence of curvature is based on a set of assumptions.  One assumption
is that the electron spectrum has the form of equation~(\ref{eqn2}).  This
functional form excludes the effects of a synchrotron cooling break
\citep{vol05}.  If the magnetic field strength is as large as 150~$\mu$G,
then synchrotron cooling may significantly steepen the highest-energy end of
the electron spectrum \citep{kse05a}. Since the fits are sensitive only to
the net amount of curvature between radio and X-ray frequencies, inclusion
of a cooling break would require a larger amount of curvature (\ie, more
flattening than the amount of curvature reported here) to compensate for the
spectral steepening of the cooling break. Another assumption is that the
shape of the radio spectrum in each of the 13 regions is the same as the
shape of the radio spectrum of the entire remnant.  If this assumption is
invalid, then the best-fit values of the spectral index and curvature may be
inaccurate for any given region. Yet, the mean values of the index and
curvature are insensitive to such spatial variations.
A third assumption is that the shape of the synchrotron spectrum is uniform
inside each of the 13 regions.  Some young remnants, such as Cas~A
\citep{and96} and Kepler \citep{del02}, exhibit evidence of radio spectral
variations from one region to another. Therefore, the combination of the
synchrotron spectra of several small-scale features might naturally yield a
curved composite spectrum even if the synchrotron spectrum for each feature
is just a power law. Small extraction regions were used to minimize this
problem.
A fourth assumption is that our estimates of the fraction of the 843~MHz
flux from each region are accurate. As discussed above, the size of the
beamwidth leads to some uncertainty about the proper radio fluxes to use for
each region.  Yet, even the worst-case scenario, the results obtained using
the peak radio fluxes, requires some spectral curvature.
High-quality radio and X-ray spectra for many, small, spatially resolved
features in the remnant would eliminate the need for many of these
assumptions. Unfortunately, these kinds of data are not available for \snr.

Evidence of curvature in the synchrotron spectra of supernova remnants has
been reported previously. \citet{rey92} analyzed the remnant-integrated
radio spectra of Kepler, Tycho, and \snr.  They find hints of curvature for
Tycho and Kepler.  Unfortunately, a careful study of the radio emission from
Kepler \citep{del02} reveals intrinsic variations in the radio spectral
index from region to region.  Therefore, it is not possible to dismiss the
idea that the evidence of curvature in the remnant-integrated spectrum is
the result of the combination of several power-law spectra with different
spectral indices.  \citet{jon03} minimized this problem for Cas~A by
analyzing the radio-to-infrared synchrotron spectra of small, selected
features in the remnant.  The results of their analysis indicate that the
spectra of these regions flatten with increasing energy.  We find that their
spectra can be fitted with our curved spectral model if the curvature
parameter $a = 0.06 \pm 0.01$.  It is interesting that this value is
consistent with the mean amount of curvature for \snr\ ($0.05 \pm 0.01$).
Based on an analysis of X-ray data for a small feature on the northeastern
shell of RCW~86, \citet{vin06} report evidence of curvature in the
synchrotron spectrum of this remnant.  This claim would be strengthened if
the data were fitted with a model that includes the spectral index as a free
parameter and if there were similar evidence for more than one feature in
the remnant.

The evidence of curvature suggests that the pressure exerted by nonthermal
particles is large enough to modify the structure of the shock.  The
cosmic-ray pressure is most likely dominated by nuclei, not electrons. For
example, analyses of the synchrotron spectrum of \snr\ have been used to
estimate the total energy of the nonthermal electrons in the remnant.  These
estimates, which depend on the assumed value of the mean magnetic field
strength, include $5 \times 10^{47}$~ergs (for $B = 150$~$\mu$G; Berezhko
\etal\ 2002), $9 \times 10^{47}$~ergs (for $B = 40$~$\mu$G; Allen \etal\
2001), and $7 \times 10^{48}$~ergs (for $B = 3$~$\mu$G; Dyer \etal\ 2001).
The remnant was recently observed with the H.E.S.S.\ gamma-ray telescope
\citep{aha05a}. If the upper limits on the TeV gamma-ray flux reported by
\citet{aha05a} are accurate (cf.\ Tanimori \etal\ 1998), then a combination
of these constraints and the measured synchrotron flux can be used to show
that the mean magnetic field strength must be at least 25~$\mu$G
\citep{aha05a}. In this case, the total energy in cosmic-ray electrons is
probably no larger than about $10^{48}$~ergs.  Since this energy is a small
fraction of the total internal energy ($\sim 10^{51}$~ergs), the nonthermal
electron pressure is most likely a small fraction of the ram pressure at the
shock. Yet, the total energy in cosmic-ray nuclei, which may be about 2
orders of magnitude larger than the total energy in cosmic-ray electrons,
could be large enough to modify the structure of the shock.

\subsection{Cutoff Frequency and Diffusion Coefficient}

In addition to providing evidence of curvature, fits to the synchrotron
spectral data provide a measure of the frequency at which the synchrotron
spectrum is cut off (eq.~[\ref{eqn04}]).  The results for the cutoff
critical frequency are listed in Tables~\ref{tab2}--\ref{tab5}.  A
comparison of Tables~\ref{tab2} and \ref{tab3} (and of Tables~\ref{tab4} and
\ref{tab5}) reveals that the frequencies for the curved spectra are somewhat
lower than the frequencies for the uncurved spectra because positive
curvature causes a synchrotron spectrum to flatten with increasing energy.
To compensate, the X-ray synchrotron spectrum is steepened by reducing the
cutoff frequency.  The solid and dashed black lines in Figure~\ref{fig7}
show that the two effects more or less offset one another to produce
electron (and, hence, synchrotron) spectra that have nearly the same shape
in the X-ray synchrotron--emitting band.

\citet{bam03}, \citet{rot04}, and \citet{bam08a} fitted X-ray spectra for
the bright rims of \snr\ with models that include the XSPEC synchrotron
component {\tt srcut}. Since {\tt srcut} does not include spectral
curvature, their results should be compared with the results listed in
Table~3.  To the extent that the regions used by \citet{bam03} and
\citet{rot04} are similar to the regions used here, this comparison reveals
that their cutoff frequencies are systematically higher than the frequencies
listed in Table~\ref{tab3}. This difference is due, in part, to a difference
in the spectral indices. Our best-fit spectral indices ($\Gamma =
1.98$--2.12; Table~\ref{tab3}) are lower than the indices used by Bamba
\etal\ (2003; $\Gamma = 2.14$) and Rothenflug \etal\ (2004; $\Gamma = 2.2$).
While an index of about 2.2 is appropriate for the radio
synchrotron--emitting electrons (\ie, the electrons at momenta $p \sim
1$~GeV~$c^{-1}$), an index of 2.0 seems to better describe the X-ray
synchrotron--emitting electrons (\ie, the electrons at momenta $p \sim
10$~TeV~$c^{-1}$). For example, if the electron spectrum is not curved, then
the mean index is $2.031 \pm 0.017$ (Table~\ref{tab3}). If curvature is
included, then the effective mean index at 10~TeV is $2.005 \pm 0.027$ [\ie,
$2.221 \pm 0.013 - (0.054 \pm 0.006) \log(10~{\rm TeV}/1~{\rm GeV})$;
Table~\ref{tab2}].  In both cases, the results imply that the index near the
cutoff of the electron spectrum is about 2.0, which is consistent with the
model of \citet{ber02}. Berezhko \etal\ note that the electron spectrum is
steeper at radio synchrotron--emitting momenta than at X-ray
synchrotron--emitting momenta because GeV electrons sample only a portion of
the shock while the highest-energy electrons ``see'' the full strength of
the shock.  Although the same argument applies to the results of
\citet{bam08a}, who use a spectral index $\Gamma = 2.14$, their best-fit
break frequency for the entire northeastern rim is lower than the mean value
of $\nu_{m}$ in Table~\ref{tab3} [$\bar{\nu}_{m} = (9.07 \pm 0.98) \times
10^{16}$~Hz]. If the normalization of the synchrotron spectrum is too high,
then the best-fit frequency may be reduced.  While the normalization
reported by Bamba \etal\ (2008; 7.72~Jy at 1~GHz) is approximately correct
for the entire northeastern rim, only the flux density of the portion of the
rim associated with the very narrow region in which the X-ray synchrotron
emission is produced should be used. For these reasons, we believe that the
cut-off frequencies reported here are relatively accurate.

The frequencies listed in Tables~\ref{tab2} and \ref{tab3} are plotted in
Figure~\ref{fig8} as a function of the position angles of the regions. This
angle is measured relative to the location of the center of the remnant
(Fig.~\ref{fig1}) and is expressed in degrees from north ($0\arcdeg$)
through east ($90\arcdeg$). As shown in Figure~\ref{fig8}, there may be an
azimuthal variation in the value of the cutoff frequency. A constant value
[\ie, the weighted mean value of $\left( 4.98 \pm 0.67 \right) \times
10^{16}$~Hz, 90\% confidence level uncertainties] can be excluded at the
2.5~$\sigma$ confidence level. \citet{rot04} also report evidence of an
azimuthal variation in the cutoff frequency.  Since the frequency $\nu_{m}
\propto p_{m}^{2} B$, these results imply that the cutoff momentum of the
electron spectrum and/or the magnetic field strength varies along the
northeastern rim.  Since it is not possible to independently determine both
$p_{m}$ and $B$ using the synchrotron spectral data alone, the present
analysis does not provide a means of identifying the cause of a variation in
the cutoff frequency.

As described in the Appendix, it is possible to use a fitted value
of the cutoff critical frequency to constrain the mean electron diffusion
coefficient $\bar{\kappa}$. The upper limit on $\bar{\kappa}$, as a fraction
of the mean Bohm diffusion coefficient $\bar{\kappa}_{\rm B}$, depends on
the forward-shock velocity $u_{1}$, the cutoff critical frequency $\nu_{m}$,
and a function $f$ of the compression ratio $r$ and the ratio of the
upstream to downstream magnetic field strengths $B_{1} / B_{2}$
(eqs.~[\ref{eqna05}]--[\ref{eqna08}]).%
\footnote{The corresponding upper limit on the diffusion length $l_{m}
\approx \bar{\kappa} / u_{1}$ \citep{lag83}.  Using eqs.~(\ref{eqn03}) and
(\ref{eqna04}) yields $l_{m} \propto f u_{1} \nu_{m,2}^{-1/2} B_{2}^{-3/2}$.
If $f = 0.1875$, $u_{1} = 2300$--5000~km~s$^{-1}$, $\nu_{m,2} = 10^{17}$~Hz,
and $B_{2} > 25$~$\mu$G \citep{aha05a}, then $l_{m} < 0.17$--0.36~pc (\ie,
$l_{m} / d < 16''$--34$''$ if $d = 2.2$~kpc;
Winkler \etal\ 2003).}%
\ If $u_{1} = 2300$--5000~km~s$^{-1}$ \citep{lam96,dwa98,gha02}, $\nu_{m} =
1.1 \times 10^{17}$~Hz (Table~\ref{tab2}), and $f = 0.1875$ (\ie, $r = 4$
and $B_{1}/B_{2} \approx 0$), then $\bar{\kappa} \le (4.5$--$21)
\bar{\kappa}_{\rm B}$.  If the compression ratio is 5.2 \citep{kse05a}
instead of 4, then $f = 0.155$ and $\bar{\kappa} \le (3.7$--$18)
\bar{\kappa}_{\rm B}$.  If the cutoff frequency is about $10^{18}$~Hz
\citep{rot04} instead of $10^{17}$~Hz, then $\bar{\kappa} \le (0.5$--$2.3)
\bar{\kappa}_{\rm B}$. In each of these cases, the highest-energy electrons
have diffusion coefficients nearly as small as the Bohm coefficient (\ie,
are being accelerated about as fast as possible).

Similar results for the diffusion coefficient are reported by \citet{bam03},
\citet{yam04}, and \citet{par06}.  The latter two of these three use
equations similar to equation~(\ref{eqna05}) but obtain significantly
smaller limits for the diffusion coefficient because they use much larger
cutoff critical frequencies ($9.1 \times 10^{17}$ and $2.5 \times
10^{18}$~Hz, respectively, instead of $1.1 \times 10^{17}$~Hz).
\citet{bam03} take a different approach.  They assume that the width
($\psi_{1} d$) of the part of a filament that is in the upstream region is
determined by the diffusion length ($\kappa_{1} / u_{1}$) in this region. In
this case,
\begin{eqnarray}
  \frac{\kappa_{1}}{\kappa_{\rm B,1}}
    & =
    & \left( \frac{27 e^{3}}{4 \pi m^{3} c^{4}} \right)^{1/2}
      \left( \frac{B_{1}^{3} \sin \theta_{1}}{\nu_{m,1}} \right)^{1/2}
      \psi_{1} d u_{1}
      \label{eqn08}
    \\
    & =
    & 0.00262
      \left( \frac{B_{1}}{3\ \mu{\rm G}} \right)^{3/2}
      \left( \frac{\psi_{1}}{1\ {\rm arcsec}} \right)
      \left( \frac{d}{1\ {\rm kpc}} \right)
      \left( \frac{u_{1}}{10^{3}\ {\rm km}\ {\rm s}^{-1}} \right)
      \left( \frac{\nu_{m,1}}{10^{17}\ {\rm Hz}} \right)^{-1/2},
      \label{eqn09}
\end{eqnarray}
where $\kappa_{1}$, $\kappa_{B,1}$ [$= p c / (3 e B_{1})$], $B_{1}$,
$\nu_{m,1}$, and $\psi_{1}$ are the diffusion coefficient, Bohm diffusion
coefficient, mean magnetic field strength, cutoff critical frequency, and
filament width,%
\footnote{Here $\psi_{1}$ is the actual angular width, not the observed
angular width.}
respectively, in the upstream region and $u_{1}$ is the velocity of the
forward shock.  The numerical constant in equation~(\ref{eqn09}) was
computed assuming that the value of $\sin \theta_{1}$ is $\pi/4$, the
isotropic mean value. Like equation~(\ref{eqna05}), equation~(\ref{eqn09})
is valid only at the momentum $p = p_{m}$.  If the value of $B_{1}$ is
comparable to a typical interstellar field strength (\eg, 3~$\mu$G), then,
for reasonable values of $\psi_{1}$, $d$, $u_{1}$, and $\nu_{m,1}$,
$\kappa_{1}$ is uncomfortably small compared with the Bohm diffusion
coefficient.  There are at least three explanations for this apparent
dilemma.  One explanation is that, near the shock, $B_{1}$ may be amplified
by cosmic-ray streaming \citep{luc00,bel01,bel04}. For example,
equation~(15) of \citet{bel01} yields $B_{1} = 30$~$\mu$G (see Ksenofontov
\etal\ 2005) and, hence, $\kappa_{1} \sim \kappa_{\rm B,1}$, if the upstream
mass density $\rho_{1} = 2.4 \times 10^{-25}$~g~cm$^{-3}$ (\ie, $n_{1} =
0.1$~cm$^{-3}$), the shock velocity $u_{1} = 2300$--5000~km~s$^{-1}$, and
the cosmic-ray pressure is about 30\%--14\% of the ram pressure,
respectively. Another explanation, which is mentioned by \cite{yam04}, is
that the rate of acceleration can exceed the Bohm limit (\ie, $\kappa_{1}$
can be less than $\kappa_{\rm B,1}$) if the magnetic field is parallel to
the shock (see Jokipii 1987). A third explanation is that the width of a
filament represents the size of the region in which the magnetic field is
strong \citep{poh05} instead of the diffusion length.  In this case,
equations~(\ref{eqn08}) and (\ref{eqn09}) are not useful because an
assumption upon which they are based is not valid.

Independent of the technique used to estimate or constrain the diffusion
length, it seems likely that the highest-energy electrons in \snr\ are
diffusing close to the Bohm limit. Otherwise, the rate of acceleration is
too low for electrons to reach energies high enough to produce the observed
X-ray synchrotron radiation. Yet, it is important to note that the limit on
the diffusion coefficient described in the Appendix (and the value obtained
using eqs.~[\ref{eqn08}] and [\ref{eqn09}]) is only an estimate, for the
following reasons:

1. The fitted values of the cutoff frequency are sensitive to the assumed
shape of the electron spectrum at momenta near the cutoff in the electron
spectrum.  The actual functional form of the electron spectrum may be
significantly more complicated than equation~(\ref{eqn2}). As an example of
the extent to which differences in the shape of the electron spectrum can
affect the limit, consider the results obtained using the models with and
without curvature. If the curved model is used, then the maximum cut-off
critical frequency is $1.1 \times 10^{17}$~Hz (Table~\ref{tab2}) and
$\bar{\kappa} \le (4.5$--$21) \bar{\kappa}_{\rm B}$ (for $f = 0.1875$,
$B_{1}/B_{2} \approx 0$, and $u_{1} = 2300$--5000~km~s$^{-1}$).  If the
model does not include curvature, then $\nu_{m} = 3.0 \times 10^{17}$~Hz
(Table~\ref{tab3}) and $\bar{\kappa} \le (1.7$--$7.8) \bar{\kappa}_{\rm B}$.

2. There is evidence that the cutoff frequency varies from one region to
another in \snr\ (Fig.~\ref{fig8}; Rothenflug \etal\ 2004). Here the
frequency used to compute the limit is the largest value in
Table~\ref{tab2}.  Therefore, our constraint on the diffusion coefficient of
\snr\ applies only to the region where the cutoff frequency is largest. The
limits for other regions of the remnant are larger.

3. The limit is computed assuming that the electron momentum $p = p_{m}$.
Therefore, the limit only applies to electrons at the cutoff momentum [\ie,
$\bar{\kappa} = \bar{\kappa}(p = p_{m})$].  An advantage to using this
momentum is that the limit does not depend on the functional form of
$\bar{\kappa}(p)$.

4. The limit is based on the assumption that the cutoff is due to
synchrotron losses.  If some other process, such as the escape of electrons
from the acceleration region or the length of time over which particles are
accelerated, determines the cutoff momentum, then the diffusion coefficient
is lower than the right-hand side of equation~(\ref{eqna05}).  If this
equation yields a limit close to 1, then the results indicate that
synchrotron losses are important for electrons that have momenta $p \ge
p_{m}$.

5. The value of the function $f$ depends on the unknown compression ratio
and the unknown ratio of the upstream to downstream magnetic field strengths
(eq.~[\ref{eqna07}]).  A value of $f = 0.1875$ (\ie, $r = 4$ and $B_{1} /
B_{2} \approx 0$) is used here because the compression ratio is probably at
least as large as 4 \citep{blo01,kse05a,war05,cas08a}.  If $r > 4$, then $f
< 0.1875$ and our upper limit on the diffusion coefficient is overestimated.
Likewise, the upper limit is overestimated if $B_{1} / B_{2}$ is
significantly larger than zero.

6. Equation~(\ref{eqna05}) was derived assuming an isotropic pitch-angle
distribution.  While such a distribution is expected if electrons diffuse in
the Bohm limit, the actual distribution may be different.



\section{Conclusions}
\label{con}

We have performed a joint spectral analysis of some {\sl Chandra} ACIS X-ray
data and MOST radio data for 13 small regions along the bright northeastern
rim of the supernova remnant \snr.  The data were fitted with a model that
includes a synchrotron emission component. This component is based on an
electron spectrum that has a momentum-dependent spectral index. The rate of
change in the index for each decade in momentum is a free parameter of the
fit. If the assumptions described in \S\ \ref{curv} are valid, then the
results of the spectral analysis, which show that the synchrotron spectra of
\snr\ are curved, can be interpreted as evidence of curvature in the
GeV-to-TeV electron spectra. The mean amount of curvature in the electron
spectra is qualitatively consistent with predictions of the amount of
curvature in the proton spectrum of \snr\ \citep{ell00}. The best-fit
power-law index at 1~GeV (\ie, at radio synchrotron--emitting momenta) is
$2.221^{+0.013}_{-0.012}$. Including the effect of curvature, the effective
spectral index at about 10~TeV (\ie, at X-ray synchrotron--emitting momenta)
is $2.005 \pm 0.027$ (90\% confidence level uncertainties).  This effective
index is consistent with the predictions of \citet{ber02}.

The evidence of curved electron spectra suggests that cosmic rays are not
``test'' particles.  The cosmic-ray pressure at the shock is large enough to
modify the structure of the shock.  Since nonthermal electrons contain only
about 0.1\% (\ie, $10^{48}$~ergs) or less of the total internal energy, the
results provide indirect evidence of a much more energetic population of
cosmic-ray protons.  Collectively, the evidence of (1) spectral curvature in
Cas~A \citep{jon03}, RCW~86 \citep{vin06}, and \snr, (2) an unusually low
electron temperature in 1E~0102.2$-$7219 \citep{hug00}, and (3) a
compression ratio greater than 4 in Tycho \citep{war05} and \snr\
\citep{cas08a} suggests that efficient particle acceleration may be a common
feature of young, shell-type supernova remnants.

The results of the spectral analysis also determine the ``cutoff critical
frequency'' $\nu_{m}$.  This frequency seems to vary from one region to
another (see also Rothenflug \etal\ 2004), which implies that the
exponential cutoff momentum of the electron spectrum and/or the strength of
the magnetic field varies.  It is not possible to identify the cause of the
variation using the synchrotron spectral data alone.

As described in the Appendix, the cutoff frequency can be used to set an
upper limit on the mean diffusion coefficient $\bar{\kappa}$ of the
highest-energy electrons.  Aside from the cutoff frequency, this limit
depends (strongly) on the velocity of the forward shock $u_{1}$ and (weakly)
on the compression ratio $r$ and the ratio of the upstream to downstream
magnetic field strengths $B_{1} / B_{2}$.  If $\nu_{m} = 1.1 \times
10^{17}$~Hz, $u_{1} = 2300$--5000~km~s$^{-1}$, $r = 4$, and $B_{1} / B_{2}
\approx 0$, then $\bar{\kappa} < (4.5$--$21) \bar{\kappa}_{\rm B}$, where
$\bar{\kappa}_{\rm B}$ is the mean Bohm diffusion coefficient.  This result
implies that at least some of the highest-energy electrons in \snr\ diffuse
close to the Bohm limit (\ie, are accelerated about as fast as possible),
which provides additional support for the idea that Galactic cosmic rays are
predominantly accelerated by the shocks of supernova remnants.



\acknowledgments

We gratefully acknowledge the guidance of Don Ellison.  This work grew out
of discussions with him regarding the shape of cosmic-ray spectra.  We thank
Chuck Dermer, whose encouragement led us to develop a synchrotron model for
a curved electron spectrum.  This work benefited substantially from
discussions with Tom Jones and Steve Reynolds and from comments by the
anonymous referee. G.\ E.\ A.\ and J.\ C.\ H.\ are supported by contract
SV3-73016 between MIT and the Smithsonian Astrophysical Observatory.  The
Chandra X-Ray Center at the Smithsonian Astrophysical Observatory is
operated on behalf of NASA under contract NAS8-03060.



\appendix

\section{Diffusion coefficient}
\label{appa}

This appendix describes how measurements of or inferences about the shock
velocity and cutoff frequency can be used to place an upper limit on the
electron diffusion coefficient.  Since the mean rate of synchrotron losses
cannot exceed the mean rate of energy gains at momenta below the cutoff of
the electron spectrum,
\begin{equation}
  \left( \frac{dE}{dt} \right)_{\rm acc}
    \ge \
    - \left( \frac{dE}{dt} \right)_{\rm sync}
  \label{eqna01}
\end{equation}
in this range.  The rate of acceleration at $p = p_{m} \gg mc$ is given
by
\begin{equation}
  \left( \frac{dE}{dt} \right)_{\rm acc}
    =
    \frac{p_{m} c}{3} \,
    \left( \frac{\kappa_{1}}{u_{1}} + \frac{\kappa_{2}}{u_{2}} \right)^{-1}
    \left( u_{1} - u_{2} \right)
  \label{eqna02}
\end{equation}
(Lagage \& Cesarsky 1983).  If the rates of synchrotron losses in the
upstream and downstream regions are weighted by the mean residence times in
these regions [$\Delta t_{1} = 4 \kappa_{1} / (u_{1} v)$ and $\Delta t_{2} =
4 \kappa_{2} / (u_{2} v)$; Webb \etal\ 1984], then
\begin{equation}
  - \left( \frac{dE}{dt} \right)_{\rm sync}
    =
    \frac{e^{4}}{6 \pi \epsilon_{0} m^{4} c^{3}} \,
    p_{m}^{2}
    \left( \frac{\kappa_{1}}{u_{1}} B_{1}^{2} \sin^{2} \theta_{1} +
      \frac{\kappa_{2}}{u_{2}} B_{2}^{2} \sin^{2} \theta_{2} \right)
    \left( \frac{\kappa_{1}}{u_{1}} + \frac{\kappa_{2}}{u_{2}}\right)^{-1}
  \label{eqna03}
\end{equation}
(Blumenthal \& Gould 1970), where $\kappa_{1}$ and $\kappa_{2}$ are the
upstream and downstream electron diffusion coefficients perpendicular to the
shock, $u_{1}$ and $u_{2}$ are the speeds of the upstream and downstream
material relative to the shock, $v$ is the velocity of a particle,
$\epsilon_{0}$ is the permittivity of free space, $B_{1}$ and $B_{2}$ are
the mean upstream and downstream magnetic field strengths, and $\theta_{1}$
and $\theta_{2}$ are the pitch angles between the electron momentum and
magnetic field vectors in the upstream and downstream regions, respectively.
If $\sin \theta$ in equation~(\ref{eqn03}) is replaced by the isotropic mean
value of $\pi / 4$ and if $\sin^{2} \theta_{1}$ and $\sin^{2} \theta_{2}$ in
equation~(\ref{eqna03}) are replaced by the isotropic mean value of
$\case{2}{3}$, then a combination of equations~(\ref{eqn03}),
(\ref{eqna01}), (\ref{eqna02}), and (\ref{eqna03}) yields
\begin{eqnarray}
  \frac{\bar{\kappa}}{\bar{\kappa}_{\rm B}}
    & \le
    & \frac{27 \pi \epsilon_{0} m c}{16 e^{2}}
      \frac{f u_{1}^{2}}{\nu_{m,2}}
    \label{eqna04}
    \\
    & \le
    & 0.936
      \left( \frac{f}{0.1875} \right)
      \left( \frac{u_{1}}{10^{3}\ {\rm km}\ {\rm s}^{-1}} \right)^{2}
      \left( \frac{\nu_{m,2}}{10^{17}\ {\rm Hz}} \right)^{-1},
    \label{eqna05}
\end{eqnarray}
where
\begin{equation}
\bar{\kappa}
  =
  \left( \frac{B_{1}^{2}}{u_{1}} \kappa_{1} +
    \frac{B_{2}^{2}}{u_{2}} \kappa_{2} \right)
  \left( \frac{B_{1}^{2}}{u_{1}} + \frac{B_{2}^{2}}{u_{2}} \right)^{-1},
  \label{eqna06}
\end{equation}
\begin{equation}
\bar{\kappa}_{\rm B}
  =
  \left( \frac{B_{1}^{2}}{u_{1}} \kappa_{\rm B,1} +
    \frac{B_{2}^{2}}{u_{2}} \kappa_{\rm B,2} \right)
  \left( \frac{B_{1}^{2}}{u_{1}} + \frac{B_{2}^{2}}{u_{2}} \right)^{-1},
  \label{eqna07}
\end{equation}
the upstream and downstream Bohm diffusion coefficients $\kappa_{\rm B,1}$
and $\kappa_{\rm B,2}$ are equal to $p_{m} c / (3 e B_{1})$ and $p_{m} c /
(3 e B_{2})$, respectively, at a momentum $p = p_{m}$, $\nu_{m,2}$ is the
critical frequency of an electron with this momentum in a magnetic field $B
= B_{\rm 2}$,
\begin{equation}
  f
    =
    \frac{r - 1}{r \left[ r + \left( B_{1} / B_{2} \right) \right]},
  \label{eqna08}
\end{equation}
and the compression ratio $r = u_{1} / u_{2}$.  The value of 0.1875 in
equation~(\ref{eqna05}) was calculated assuming $r = 4$ and $B_{1} / B_{2}
\approx 0$.  If the compression ratio is greater than 4 or $B_{1} / B_{2}$
is significantly larger than zero, then $f < 0.1875$.  In no case can $f$ be
larger than 0.25.  If $B_{1} = B_{2}$, then equation~(\ref{eqna05}) is
identical to the analogous equation in \cite{sta06}. 
Equation~(\ref{eqna05}) is similar to equation~(22) of \cite{aha99a},
equation~(12) of \cite{laz04}, equation~(A.4) of \cite{yam04}, and
equation~(22) of \cite{par06}, except these authors use the peak frequency
instead of the critical frequency.  The latter three also use $\sin \theta =
1$ instead of $\pi / 4$.





\plotone{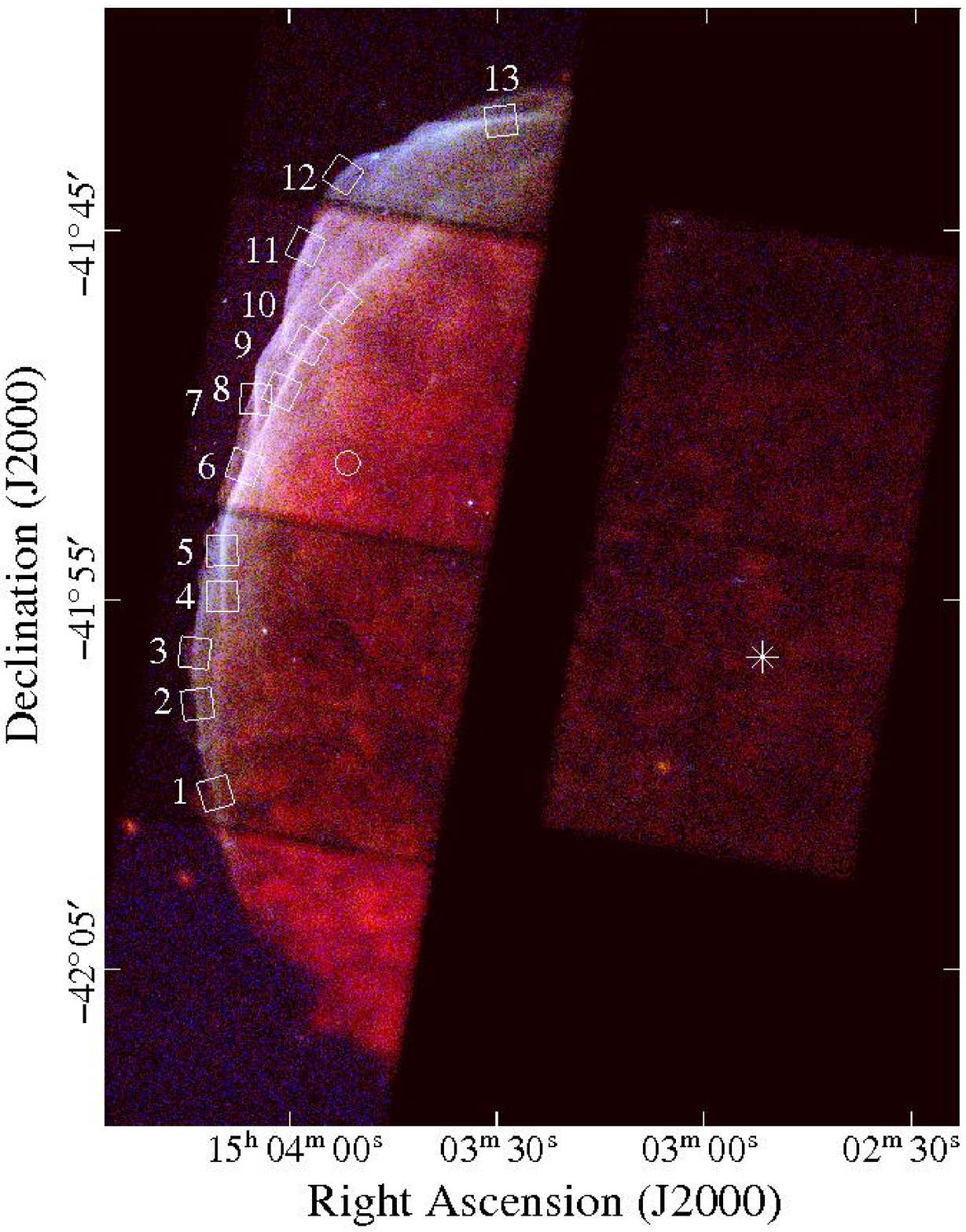}
\figcaption[f1b.eps]{
Color-coded ACIS image of the northeastern rim of \snr. Red, green, and blue
correspond to the energy bands 0.4--1, 1--2, and 2--7~keV, respectively.
Most of the emission at energies greater than 1~keV is concentrated in the
whitish filaments along the rim.  The edges of the six square CCDs used to
observe the remnant are evident. The relatively high brightnesses of the CCD
with the circle on it and the CCD in the lower left corner are due to
enhanced low-energy sensitivities for these two devices. The circle at
$\alpha = 15^{\rm h} 03^{\rm m} 51.56^{\rm s}$ and $\delta = -41\arcdeg
51\arcmin 18.8\arcsec$ (J2000) indicates the location of the nominal aim
point of the telescope. The asterisk at $\alpha = 15^{\rm h} 02^{\rm m}
51.7^{\rm s}$ and $\delta = -41\arcdeg 56\arcmin 33.0\arcsec$ is the
location of the center of the supernova remnant \citep{win97}.  The 13
squares are the X-ray spectral extraction regions.
\label{fig1}}


\newpage

\plotone{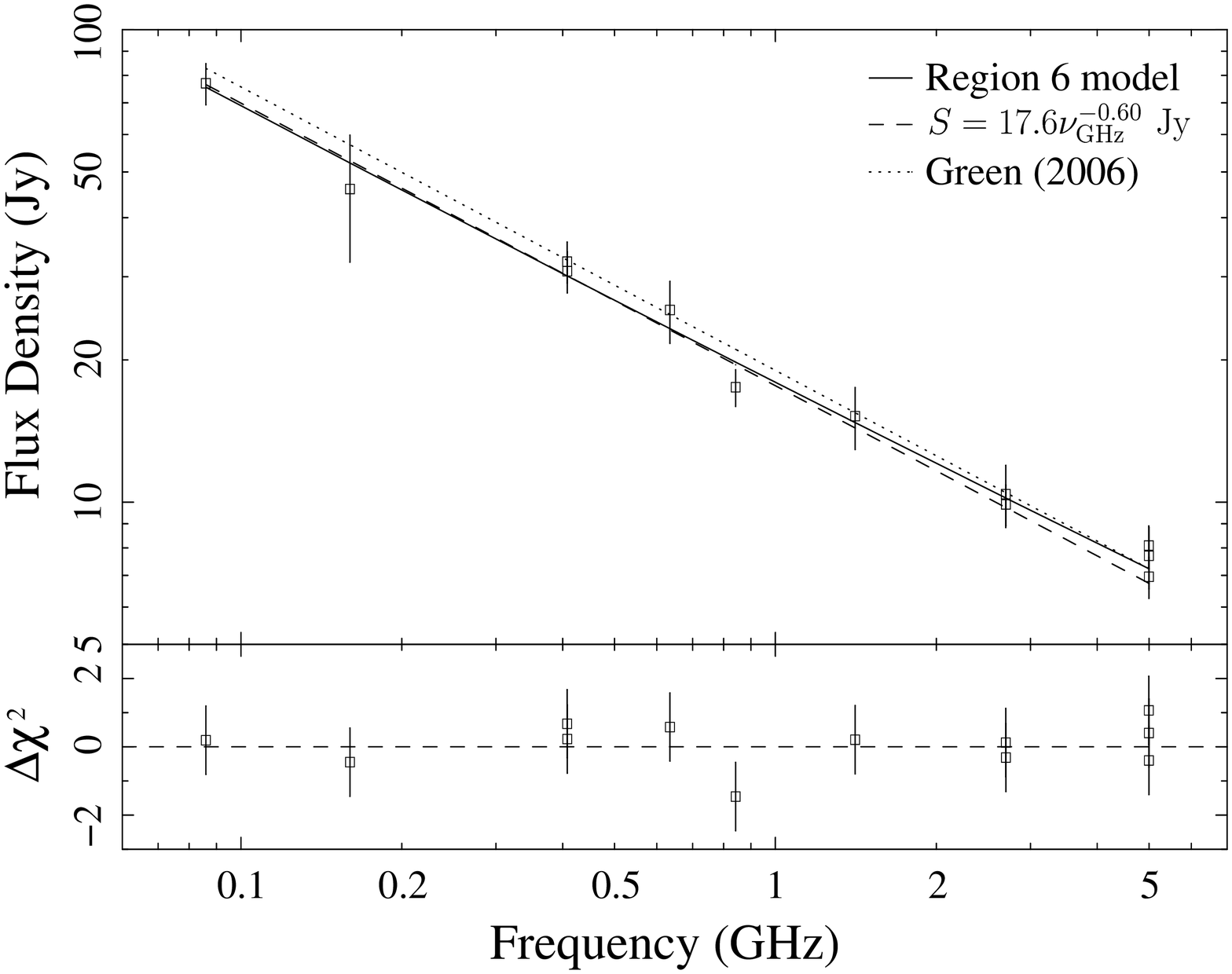}
\figcaption[f2.eps]{
Radio spectrum for the entire supernova remnant \snr.  {\sl Top}: The data
points (Table~\ref{tab1}) and three models. The dashed line is the best-fit
power law, $S(\nu) = 17.6^{+6.3}_{-4.8}[\nu/(1~{\rm GHz})]^{-\alpha}$~Jy,
where $\alpha = 0.60^{+0.08}_{-0.09}$.  The dotted line is a power law using
Green's (2006) spectral parameters [$S(\nu) = 19 [\nu/(1~{\rm
GHz})]^{-0.6}$~Jy]. The solid line is the best-fit (curved) synchrotron
model for region~6 divided by the factor 0.00175 (see Table~\ref{tab2}).
{\sl Bottom}: Differences between the data points and the solid line,
divided by the uncertainties in the data points.
\label{fig2}}


\plotone{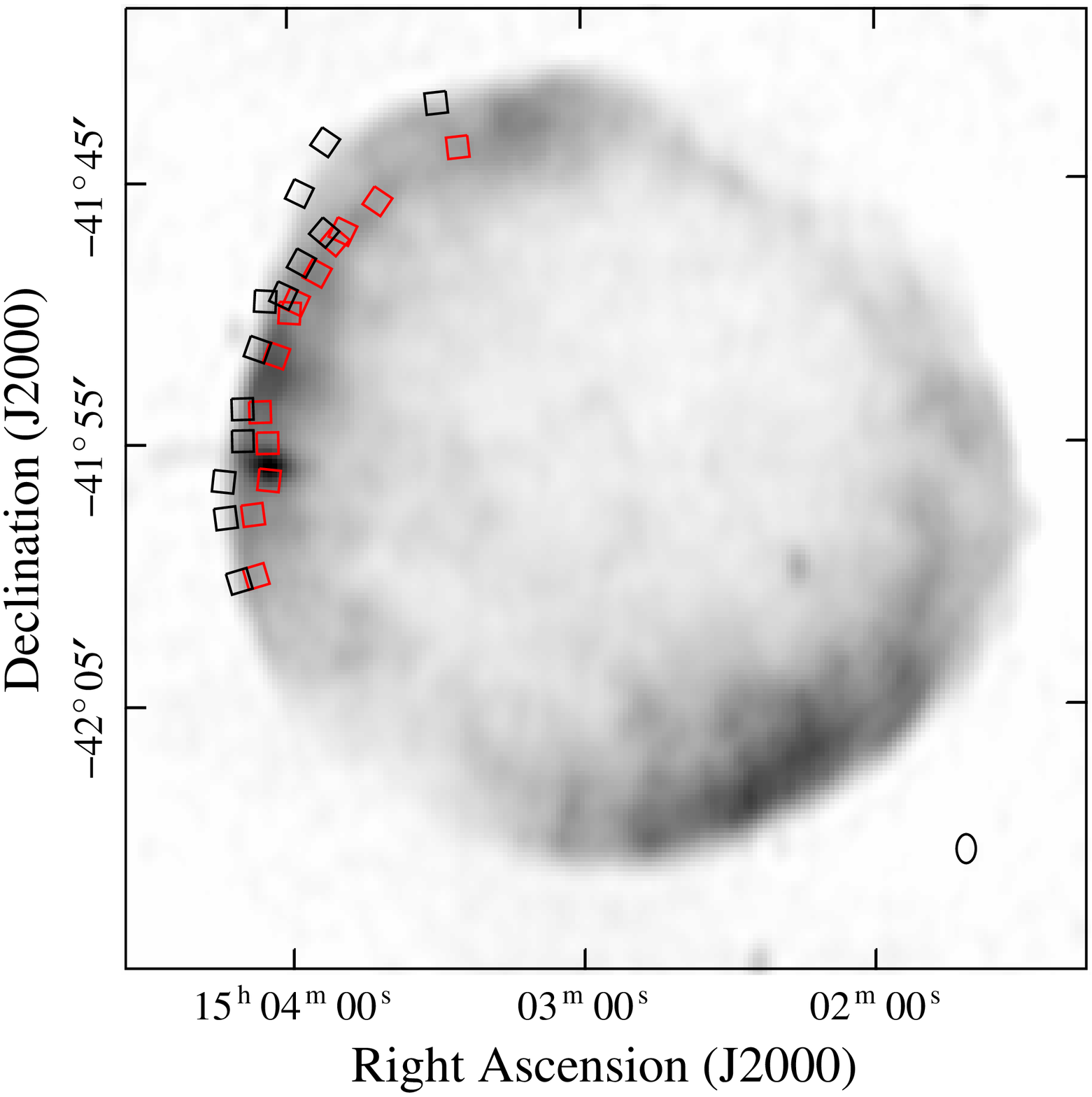}
\figcaption[f3.eps]{
An 843~MHz MOST image of \snr, courtesy of R.\ Roger \citep{rog88}. The
ellipse at lower right indicates the half-power beamwidth.  The 13 black
squares are the locations of the X-ray spectral extraction regions.  These
regions were used to obtain the ``cospatial'' radio fluxes
(Tables~\ref{tab2} and \ref{tab3}). The red squares are the locations used
to obtain the peak radio fluxes (Tables~\ref{tab4} and \ref{tab5}). The
ratios of the flux densities of the black and red regions to the total flux
density are listed in the $\zeta$ columns of Tables~\ref{tab2}--\ref{tab5}.
The bright spot at about $\alpha = 15^{\rm h} 04^{\rm m} 04^{s}$ and $\delta
= -41\arcdeg 55\arcmin 48\arcsec$ is produced by an extragalactic source
\citep{rey86}.
\label{fig3}}


\plotone{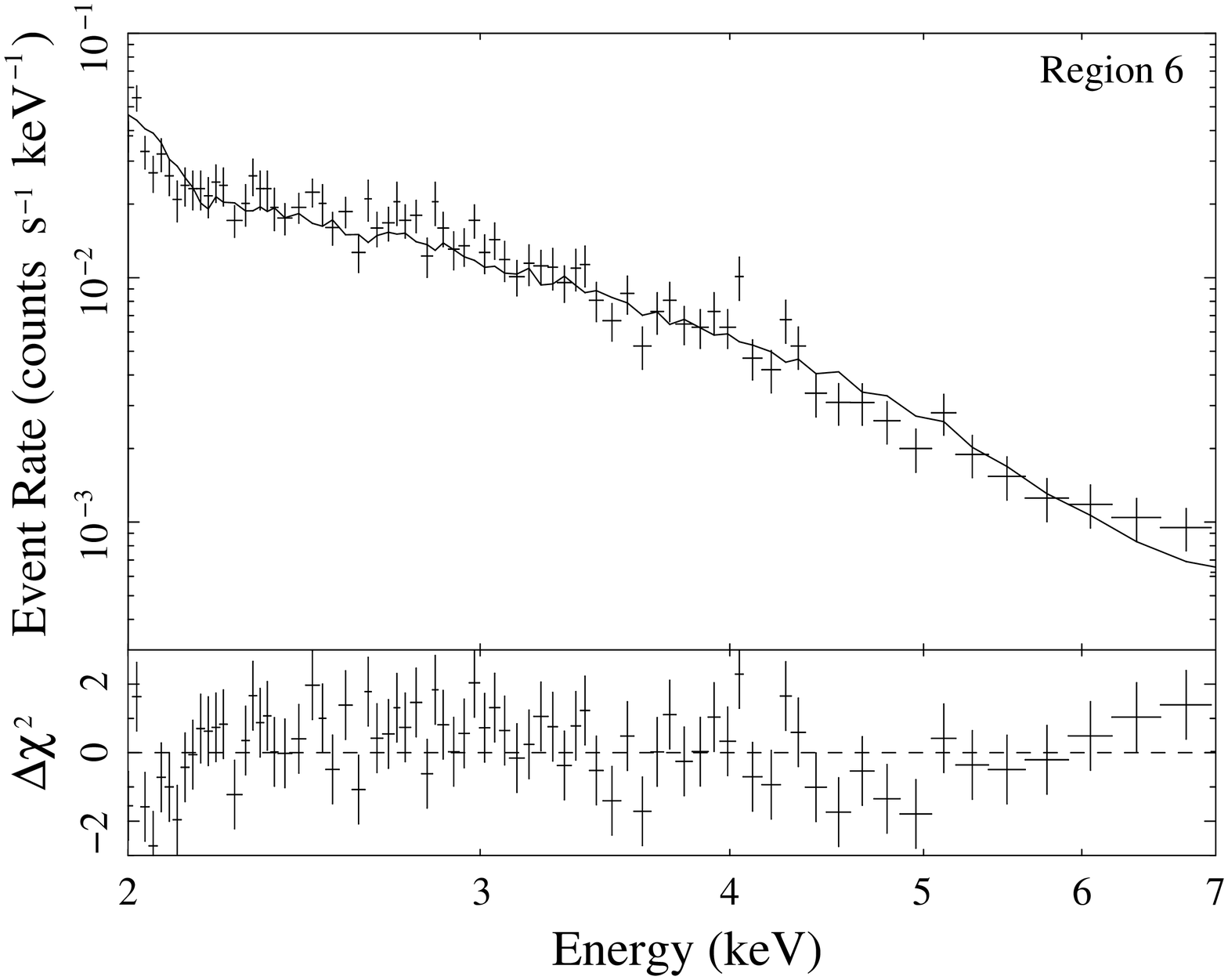}
\figcaption[f4.eps]{
ACIS spectrum for region~6 (see Fig.~\ref{fig1}) of the bright northeastern
rim of \snr.  {\sl Top}: Sum of the source and background spectra ({\sl data
points}) and the sum of the best-fit model and background spectra ({\sl
solid line}).  The model displayed here includes spectral curvature as a
free parameter (see Table~\ref{tab2}).  {\sl Bottom}: Differences between
the data points and the solid line, divided by the uncertainties in the data
points.
\label{fig4}}


\plotone{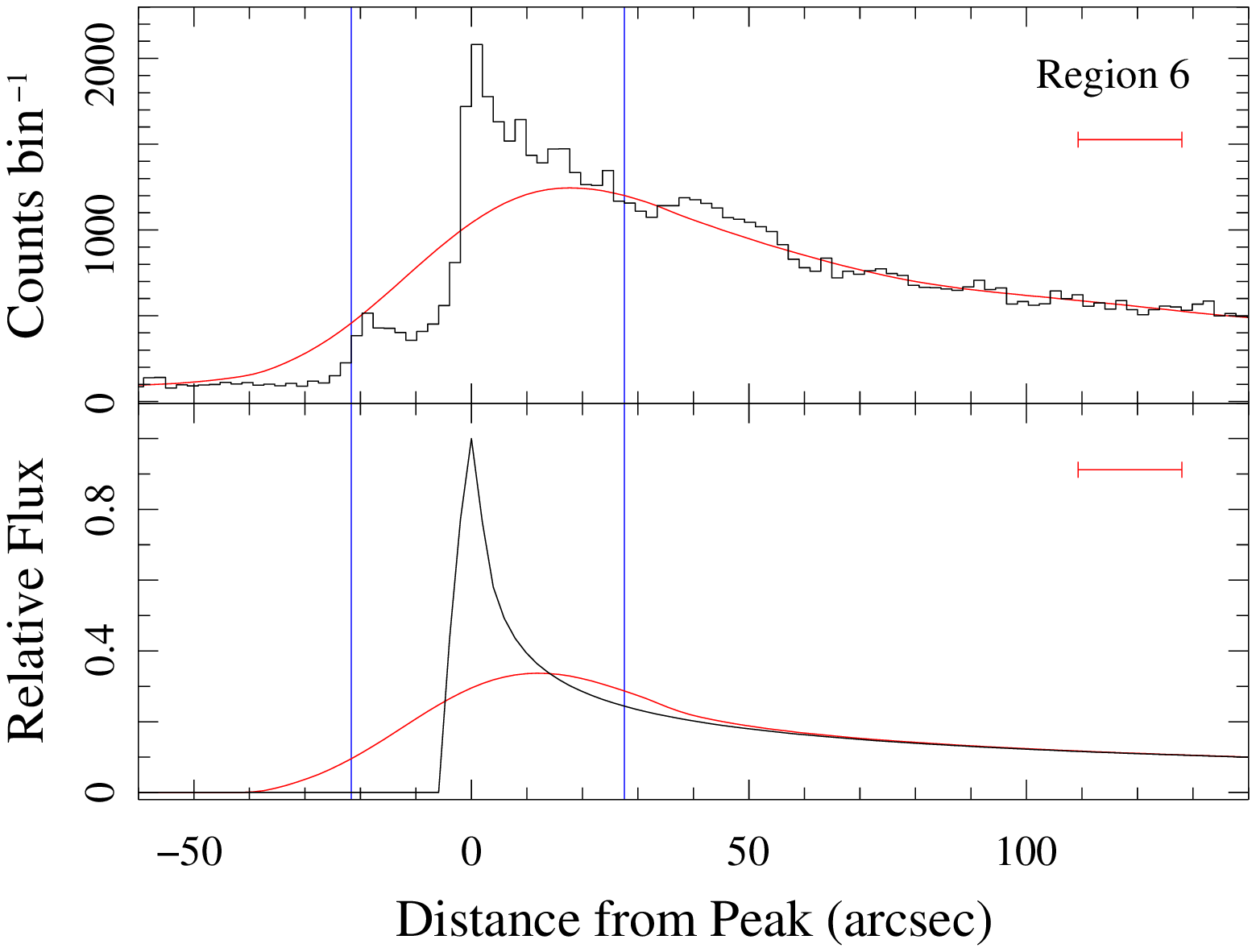}
\figcaption[f5.eps]{
X-ray emission profiles along a $49 \arcsec$-wide strip passing through
region~6 toward the center of \snr. The black histogram in the top panel is
the 2--7~keV ACIS data. For comparison, the black line in the bottom panel
is a model based on the assumption that the emission is produced in a
uniformly emitting shell that is $6 \arcsec$ wide. The red curves are the
data and model smoothed to the resolution of the MOST image. (The horizontal
red bars are the 1~$\sigma$ widths of the Gaussian smoothing function.) The
reduction in spatial resolution causes the peak to shift downstream, but the
offset is small compared with the width of the extraction region. (The blue
lines are the boundaries of region~6.) Therefore, it seems unlikely that the
offsets between the X-ray and radio peaks (see Fig.~\ref{fig3} and
Tables~\ref{tab4} and \ref{tab5}) are due to the different X-ray and radio
resolutions.
\label{fig5}}


\plotone{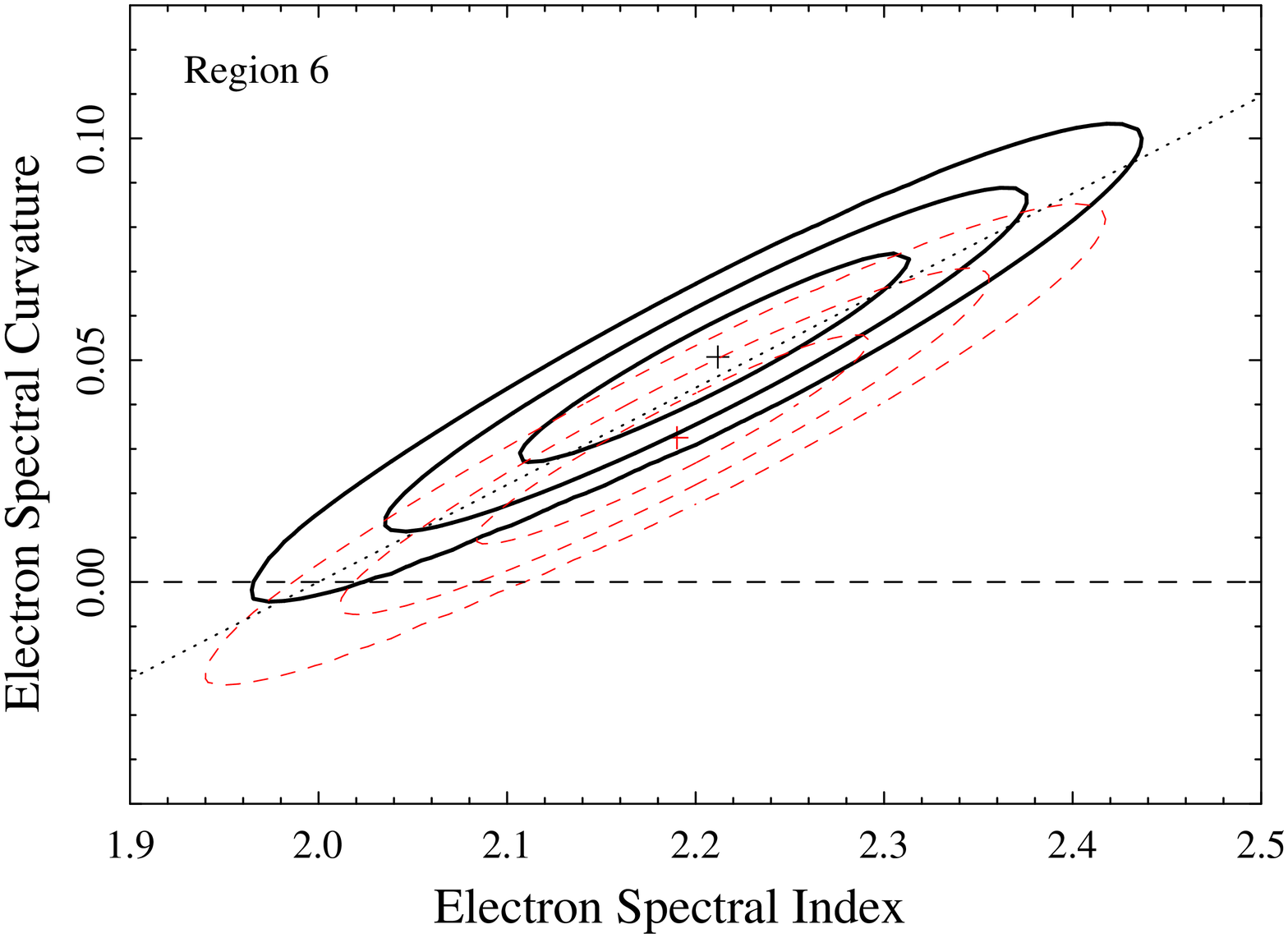}
\figcaption[f6.eps]{
The 1, 2, and 3~$\sigma$ confidence contours for region~6 (see
Fig.~\ref{fig1}) in the parameter space defined by the electron spectral
index $\Gamma$ and electron spectral curvature $a$ (see eq.~[\ref{eqn2}]).
The solid black and dashed red contours are the results obtained using the
cospatial (Table~\ref{tab2}) and peak (Table~\ref{tab4}) radio fluxes,
respectively. The plus signs indicate the best-fit values of the index and
curvature.  The dotted black line is the expected relationship between
$\Gamma$ and $a$ (eq.~[\ref{eqn07}]). The dashed black line is the line
along which the electron spectrum is not curved (\ie, $a = 0$). For
region~6, a positive curvature (\ie, a result above the dashed line) is
favored at about the 2.7 and 1.6~$\sigma$ confidence levels for the black
and red contours, respectively.
%
\label{fig6}}


\plotone{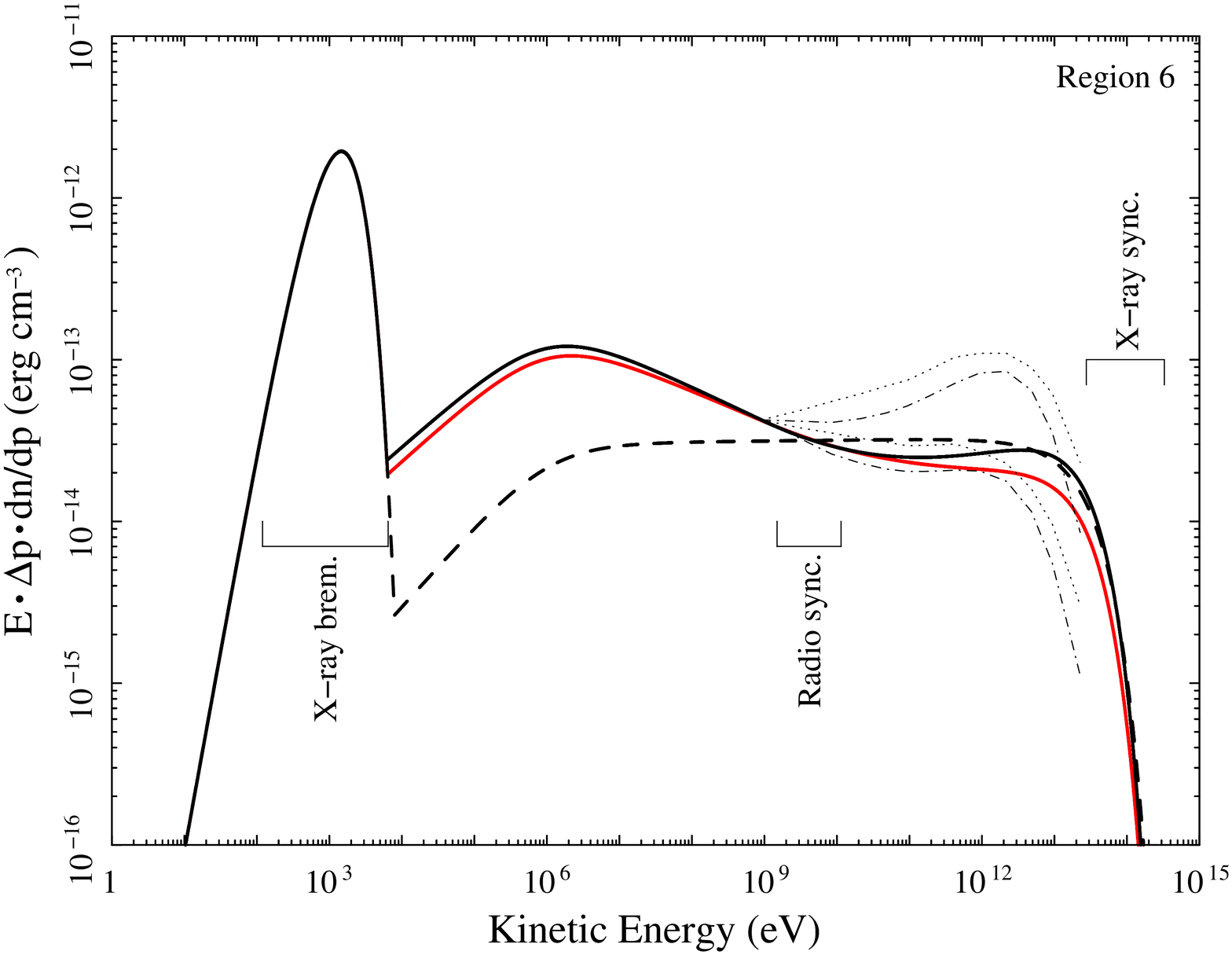}
\figcaption[f7.eps]{
Best-fit electron number density spectra for region 6 (see Fig.~\ref{fig1}).
The solid and dashed black lines are the curved (Table~\ref{tab2}) and
uncurved (Table~\ref{tab3}) models, respectively, using the cospatial radio
flux.  The solid red line is the curved model using the peak radio flux
(Table~\ref{tab4}). The upper dotted and dot-dashed curves are the models
for the proton (not electron) spectrum of \snr\ presented by \cite{ell00}.
These lines are plotted only for momenta $p > mc$ (\ie, the range of momenta
to which our fits are sensitive) and are normalized to the solid black line
at a kinetic energy of 0.9~GeV.  As described in the text, the lower dotted
and dot-dashed curves are the same pair of models multiplied by $[E / (0.9\
{\rm GeV})]^{-0.2}$.  The amounts of spectral curvature in the solid lines
are consistent with the amounts of curvature in the lower dotted and
dot-dashed curves. From left to right, the three bracketed energy bands
contain the electrons that are primarily responsible for the observed
thermal bremsstrahlung, radio synchrotron, and X-ray synchrotron emission.
\label{fig7}}


\plotone{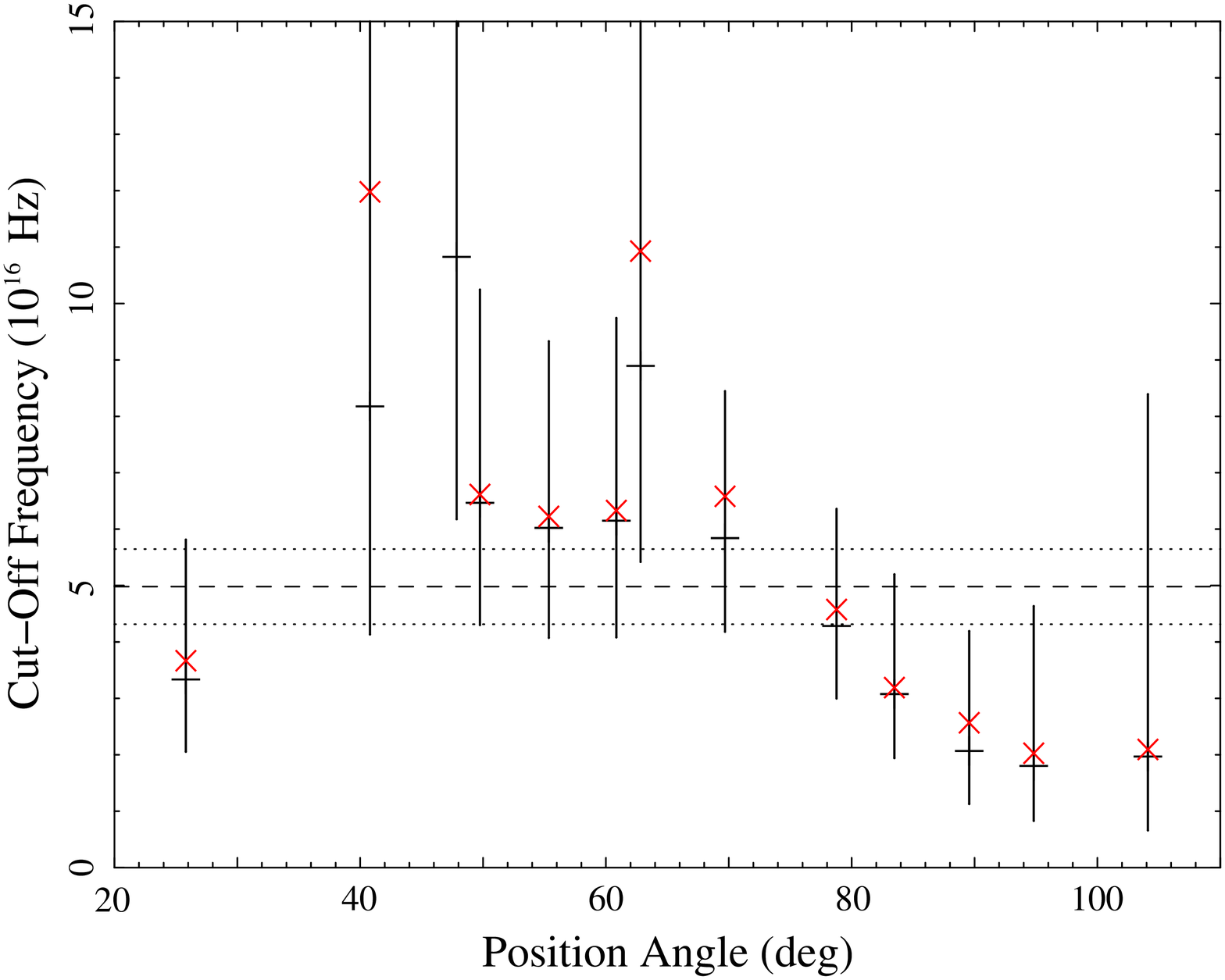}
\figcaption[f8.eps]{
Cutoff frequency as a function of position angle (measured counterclockwise
from north through east).  From right to left, the 13 pairs of points are
for regions 1--13, respectively (see Fig.~\ref{fig1}). The black data points
and 90\% confidence intervals are the results obtained using a curved
electron spectrum and the cospatial radio fluxes (Table~\ref{tab2}).  The
red data points, which have similar confidence intervals, are the results
obtained using the peak radio fluxes (Table~\ref{tab4}). The dashed line is
the weighted mean value ($4.98 \times 10^{16}$~Hz) of the black points.  The
corresponding 90\% confidence level interval [(4.31--$5.65) \times
10^{16}$~Hz] lies between the two dotted lines. A constant value for the
cutoff frequency can be excluded at the 2.5~$\sigma$ confidence level.
\label{fig8}}



\clearpage
\input{tab1.tex}

\clearpage
\input{tab2.tex}

\clearpage
\input{tab3.tex}

\clearpage
\input{tab4.tex}

\clearpage
\input{tab5.tex}



\end {document}

%% file: tab1.tex
\begin{deluxetable}{cccc}
\tablecaption{Radio Data\label{tab1}}
\tablewidth{0pt}
\tablehead{
$\nu$
  & $S$
  & $\Delta S$
  & 
  \\
(GHz)
  & (Jy)
  & (Jy)
  & Refs.
}
\startdata
0.086
  & 77
  & 7.7\tablenotemark{a}
  & 1
  \\
0.16
  & 46
  & 13.8
  & 1
  \\
0.408
  & 32.3
  & 3.2
  & 1
  \\
0.408
  & 30.8
  & 3.1\tablenotemark{a}
  & 2
  \\
0.635
  & 25.5
  & 3.8
  & 1
  \\
0.843
  & 17.5
  & 1.5
  & 3
  \\
1.41
  & 15.2
  & 2.3
  & 1
  \\
2.7
  & 10.4 
  & 1.6
  & 1
  \\
2.7
  & 9.9
  & 1.0\tablenotemark{a}
  & 1, 4
  \\
5.0
  & 7.7
  & 1.2
  & 1
  \\
5.0
  & 7.0
  & 0.7\tablenotemark{a}
  & 5
  \\
5.0
  & 8.1
  & 0.8\tablenotemark{a}
  & 6
  \\
\enddata

\tablenotetext{\,}{Note.---The table lists the frequencies $\nu$, flux
densities $S$, and 1~$\sigma$ flux density uncertainties $\Delta S$ used for
the spectral analysis.  The quantities $S$ and $\Delta S$ are for the
emission from the entire supernova remnant.}

\tablenotetext{a}{Since no uncertainty was reported, it is assumed to be
10\% of $S$. The results of our spectral fits are insensitive to the assumed
fractional uncertainty at least for fractions between 5\% and 20\%.}

\tablenotetext{\ }{References.---(1) Milne 1971; (2) Stephenson \etal\ 1977;
(3) Roger \etal\ 1988; (4) Gardner \& Milne 1965; (5) Kundu 1970; (6) Milne
\& Dickel 1975.}

\end{deluxetable}

%% file: tab2.tex
\begin{deluxetable}{ccccccccccc}
\rotate
\tablecaption{Best-Fit Parameters Using the Curved Model and ``Cospatial''
  Radio Fluxes\label{tab2}}
\tablewidth{0pt}
\tablehead{
\
  & $\alpha_{\rm X}$
  & $\delta_{\rm X}$
  & $\phi$
  & $\zeta$
  &
  &
  & $\nu_{\rm m}$
  &
  \\
Region
  & (J2000)
  & (J2000)
  & (deg)
  & ($\times 10^{-4}$)
  & $\Gamma$\tablenotemark{a}
  & $a$
  & ($10^{16}$~Hz)
  & $\chi^{2}/{\rm dof}$
  & $P_{\Delta \chi^{2}}$\tablenotemark{b}
}
\startdata
1
  & 15 04 10.7
  & $-$42 00 15
  & 104.1
  & 10
  & $2.24^{+0.16}_{-0.16}$
  & $0.050^{+0.070}_{-0.059}$
  & $2.0^{+6.4}_{-1.3}$
  & 8.4/18
  & 0.17
  \\
2
  & 15 04 13.3
  & $-$41 57 51
  & 94.8
  & 8.9
  & $2.27^{+0.13}_{-0.15}$
  & $0.071^{+0.051}_{-0.047}$
  & $1.8^{+2.8}_{-1.0}$
  & 12.9/22
  & 0.012
  \\
3
  & 15 04 13.7
  & $-$41 56 27
  & 89.6
  & 6.9
  & $2.27^{+0.13}_{-0.13}$
  & $0.080^{+0.039}_{-0.039}$
  & $2.1^{+2.1}_{-1.0}$
  & 11.0/27
  & 0.00055
  \\
4
  & 15 04 09.7
  & $-$41 54 54
  & 83.5
  & 24
  & $2.22^{+0.12}_{-0.12}$
  & $0.051^{+0.030}_{-0.031}$
  & $3.1^{+2.1}_{-1.2}$
  & 50.2/46
  & 0.0073
  \\
5
  & 15 04 09.7
  & $-$41 53 41
  & 78.8
  & 15
  & $2.23^{+0.11}_{-0.12}$
  & $0.060^{+0.026}_{-0.027}$
  & $4.3^{+2.1}_{-1.3}$
  & 71.4/63
  & 0.00033
  \\
6
  & 15 04 06.5
  & $-$41 51 25
  & 69.7
  & 18
  & $2.21^{+0.11}_{-0.11}$
  & $0.051^{+0.025}_{-0.026}$
  & $5.8^{+2.6}_{-1.6}$
  & 91.5/84
  & 0.0014
  \\
7
  & 15 04 04.8
  & $-$41 49 35
  & 62.8
  & 11
  & $2.19^{+0.11}_{-0.11}$
  & $0.036^{+0.027}_{-0.027}$
  & $8.9^{+6.9}_{-3.5}$
  & 69.2/60
  & 0.030
  \\
8
  & 15 04 01.1
  & $-$41 49 22
  & 60.8
  & 17
  & $2.20^{+0.11}_{-0.11}$
  & $0.042^{+0.026}_{-0.027}$
  & $6.1^{+3.6}_{-2.0}$
  & 69.4/71
  & 0.011
  \\
9
  & 15 03 57.3
  & $-$41 48 08
  & 55.4
  & 19
  & $2.20^{+0.11}_{-0.11}$
  & $0.042^{+0.026}_{-0.027}$
  & $6.0^{+3.3}_{-1.9}$
  & 73.0/72
  & 0.010
  \\
10
  & 15 03 52.6
  & $-$41 46 58
  & 49.7
  & 17
  & $2.20^{+0.11}_{-0.12}$
  & $0.042^{+0.026}_{-0.027}$
  & $6.5^{+3.7}_{-2.2}$
  & 71.4/69
  & 0.011
  \\
11
  & 15 03 57.7
  & $-$41 45 28
  & 47.9
  & 3.8
  & $2.20^{+0.11}_{-0.11}$
  & $0.057^{+0.028}_{-0.028}$
  & $10.8^{+10.3}_{-4.6}$
  & 65.3/61
  & 0.00079
  \\
12
  & 15 03 52.3
  & $-$41 43 31
  & 40.8
  & 3.0
  & $2.21^{+0.11}_{-0.11}$
  & $0.060^{+0.032}_{-0.032}$
  & $8.2^{+10.9}_{-4.1}$
  & 32.2/37
  & 0.0018
  \\
13
  & 15 03 29.5
  & $-$41 42 03
  & 25.8
  & 9.5
  & $2.25^{+0.11}_{-0.12}$
  & $0.070^{+0.031}_{-0.031}$
  & $3.3^{+2.5}_{-1.2}$
  & 35.5/42
  & 0.00023
  \\
Mean
  &
  &
  &
  &
  & $2.221^{+0.013}_{-0.012}$ 
  & $0.054 \pm 0.006$
  &
  &
  &
\enddata

\tablenotetext{\,}{
Notes.---For each region, the table includes a list of the right ascension
($\alpha_{\rm X}$), declination ($\delta_{\rm X}$), and position-angle
($\phi$) coordinates of the center of the X-ray region, the 843~MHz flux
divided by the total 843~MHz flux ($\zeta$), the best-fit differential
electron spectral index $\Gamma$, curvature parameter $a$, and cutoff
frequency $\nu_{\rm m}$, the value of $\chi^{2}$ per degree of freedom
($\chi^{2} / {\rm dof}$), and the probability ($P_{\Delta \chi^{2}}$) that
the difference between the value of $\chi^{2}$ in Table~\ref{tab3} and the
value in this table is due to chance. The uncertainties, which include only
the statistical contributions, are reported at the 90\% confidence level.
These uncertainties were used to compute the weighted mean values of
$\Gamma$ and $a$. Units of right ascension are hours, minutes, and seconds,
and units of declination are degrees, arcminutes, and arcseconds.
}

\tablenotetext{a}{
As described in the text, the spectral indices are not independent from
region to region.
}

\tablenotetext{b}{
Including curvature as a free parameter significantly improves the quality
of the fit for most regions.
}

\end{deluxetable}

%% file: tab3.tex
\begin{deluxetable}{cccccccccc}
\tablecaption{Best-Fit Parameters Using the Uncurved Model and ``Cospatial''
  Radio Fluxes\label{tab3}}
\tablewidth{0pt}
\tablehead{
\
  & $\alpha_{\rm X}$
  & $\delta_{\rm X}$
  & $\phi$
  & $\zeta$
  & 
  & 
  & $\nu_{\rm m}$
  & 
  \\
Region
  & (J2000)
  & (J2000)
  & (deg)
  & ($\times 10^{-4}$)
  & $\Gamma$
  & $a$\tablenotemark{a}
  & ($10^{16}$~Hz)
  & $\chi^{2}/{\rm dof}$
}
\startdata
1
  & 15 04 10.7
  & $-$42 00 15
  & 104.1
  & 10
  & $2.12^{+0.08}_{-0.08}$
  & 0
  & $5.8^{+5.3}_{-2.4}$
  & 10.3/19
  \\
2
  & 15 04 13.3
  & $-$41 57 51
  & 94.8
  & 8.9
  & $2.08^{+0.07}_{-0.08}$
  & 0
  & $6.8^{+6.0}_{-2.8}$
  & 19.2/23
  \\
3
  & 15 04 13.7
  & $-$41 56 27
  & 89.6
  & 6.9
  & $2.03^{+0.07}_{-0.07}$
  & 0
  & $7.7^{+6.1}_{-3.0}$
  & 22.9/28
  \\
4
  & 15 04 09.7
  & $-$41 54 54
  & 83.5
  & 24
  & $2.04^{+0.05}_{-0.06}$
  & 0
  & $5.8^{+3.1}_{-1.8}$
  & 57.4/47
  \\
5
  & 15 04 09.7
  & $-$41 53 41
  & 78.8
  & 15
  & $1.98^{+0.04}_{-0.05}$
  & 0
  & $7.6^{+3.3}_{-2.1}$
  & 84.3/64
  \\
6
  & 15 04 06.5
  & $-$41 51 25
  & 69.7
  & 18
  & $1.99^{+0.04}_{-0.03}$
  & 0
  & $9.2^{+3.6}_{-2.4}$
  & 101.7/85
  \\
7
  & 15 04 04.8
  & $-$41 49 35
  & 62.8
  & 11
  & $2.05^{+0.05}_{-0.04}$
  & 0
  & $14.7^{+9.3}_{-5.2}$
  & 73.9/61
  \\
8
  & 15 04 01.1
  & $-$41 49 22
  & 60.8
  & 17
  & $2.03^{+0.04}_{-0.04}$
  & 0
  & $9.8^{+4.7}_{-3.0}$
  & 75.9/72
  \\
9
  & 15 03 57.3
  & $-$41 48 08
  & 55.4
  & 19
  & $2.03^{+0.04}_{-0.04}$
  & 0
  & $9.4^{+4.3}_{-2.7}$
  & 79.6/73
  \\
10
  & 15 03 52.6
  & $-$41 46 58
  & 49.7
  & 17
  & $2.03^{+0.04}_{-0.04}$
  & 0
  & $10.3^{+5.0}_{-3.1}$
  & 77.9/70
  \\
11
  & 15 03 57.7
  & $-$41 45 28
  & 47.9
  & 3.8
  & $2.00^{+0.04}_{-0.05}$
  & 0
  & $28.4^{+26.8}_{-12.1}$
  & 76.5/62
  \\
12
  & 15 03 52.3
  & $-$41 43 31
  & 40.8
  & 3.0
  & $2.02^{+0.06}_{-0.06}$
  & 0
  & $29.6^{+39.0}_{-14.4}$
  & 41.9/38
  \\
13
  & 15 03 29.5
  & $-$41 42 03
  & 25.8
  & 9.5
  & $2.00^{+0.05}_{-0.06}$
  & 0
  & $8.8^{+5.6}_{-3.1}$
  & 49.1/43
  \\
Mean
  &  
  &  
  &  
  &  
  & $2.031^{+0.017}_{-0.016}$
  & 0
  & 
  &  
\enddata

\tablenotetext{\,}{Notes.---See the caption for Table~\ref{tab2}. Units of
right ascension are hours, minutes, and seconds, and units of declination
are degrees, arcminutes, and arcseconds.}

\tablenotetext{a}{The curvature parameter was fixed at zero.}

\end{deluxetable}

%% file: tab4.tex
\begin{deluxetable}{ccccccccccc}
\rotate
\tablecaption{Best-Fit Parameters Using the Curved Model and Peak Radio
  Fluxes\label{tab4}}
\tablewidth{0pt}
\tablehead{
\
  & $\alpha_{R}$
  & $\delta_{R}$
  & $\Delta\Psi$
  & $\zeta$
  &
  &
  & $\nu_{m}$
  &
  \\
Region
  & (J2000)
  & (J2000)
  & (arcmin)
  & ($\times 10^{-4}$)
  & $\Gamma$\tablenotemark{a}
  & $a$
  & ($10^{16}$~Hz)
  & $\chi^{2}/{\rm dof}$
  & $P_{\Delta \chi^{2}}$\tablenotemark{b}
}
\startdata
1
  & 15 04 07.1
  & $-$42 00 05
  & 0.68
  & 17
  & $2.21^{+0.16}_{-0.16}$
  & $0.035^{+0.069}_{-0.060}$
  & $2.1^{+7.3}_{-1.4}$
  & 8.5/18
  & 0.35
  \\
2
  & 15 04 07.7
  & $-$41 57 44
  & 1.04
  & 22
  & $2.22^{+0.14}_{-0.15}$
  & $0.042^{+0.050}_{-0.047}$
  & $2.0^{+3.5}_{-1.1}$
  & 13.1/22
  & 0.15
  \\
3
  & 15 04 04.4
  & $-$41 56 25
  & 1.73
  & 30
  & $2.20^{+0.12}_{-0.14}$
  & $0.031^{+0.040}_{-0.038}$
  & $2.6^{+3.0}_{-1.3}$
  & 11.3/27
  & 0.18
  \\
4
  & 15 04 04.6
  & $-$41 55 00
  & 0.95
  & 30
  & $2.21^{+0.12}_{-0.12}$
  & $0.043^{+0.031}_{-0.030}$
  & $3.2^{+2.2}_{-1.2}$
  & 50.2/46
  & 0.021
  \\
5
  & 15 04 06.1
  & $-$41 53 48
  & 0.67
  & 22
  & $2.21^{+0.11}_{-0.12}$
  & $0.048^{+0.026}_{-0.027}$
  & $4.6^{+2.3}_{-1.4}$
  & 71.5/63
  & 0.0037
  \\
6
  & 15 04 02.6
  & $-$41 51 40
  & 0.77
  & 31
  & $2.19^{+0.11}_{-0.12}$
  & $0.033^{+0.025}_{-0.026}$
  & $6.6^{+3.1}_{-2.0}$
  & 91.8/84
  & 0.040
  \\
7
  & 15 03 59.9
  & $-$41 50 02
  & 1.03
  & 26
  & $2.16^{+0.11}_{-0.11}$
  & $0.010^{+0.027}_{-0.028}$
  & $10.9^{+9.6}_{-4.5}$
  & 69.6/60
  & 0.55
  \\
8
  & 15 03 58.6
  & $-$41 49 37
  & 0.53
  & 20
  & $2.20^{+0.11}_{-0.12}$
  & $0.037^{+0.027}_{-0.026}$
  & $6.3^{+3.8}_{-2.1}$
  & 69.5/71
  & 0.023
  \\
9
  & 15 03 54.1
  & $-$41 48 31
  & 0.72
  & 22
  & $2.20^{+0.11}_{-0.12}$
  & $0.037^{+0.026}_{-0.027}$
  & $6.2^{+3.5}_{-2.0}$
  & 73.0/72
  & 0.023
  \\
10
  & 15 03 50.5
  & $-$41 47 18
  & 0.52
  & 19
  & $2.20^{+0.11}_{-0.12}$
  & $0.038^{+0.027}_{-0.026}$
  & $6.6^{+3.9}_{-2.2}$
  & 71.4/69
  & 0.019
  \\
11
  & 15 03 47.8
  & $-$41 46 56
  & 2.35
  & 16
  & $2.16^{+0.11}_{-0.11}$
  & $0.013^{+0.027}_{-0.028}$
  & $16.0^{+19.2}_{-7.6}$
  & 65.7/61
  & 0.46
  \\
12
  & 15 03 41.7
  & $-$41 45 48
  & 3.02
  & 15
  & $2.16^{+0.11}_{-0.11}$
  & $0.010^{+0.032}_{-0.031}$
  & $12.0^{+20.7}_{-6.4}$
  & 32.6/37
  & 0.59
  \\
13
  & 15 03 25.2
  & $-$41 43 45
  & 1.88
  & 17
  & $2.22^{+0.12}_{-0.12}$
  & $0.052^{+0.031}_{-0.031}$
  & $3.7^{+2.8}_{-1.5}$
  & 35.6/42
  & 0.0062
  \\
Mean
  &
  &
  &
  &
  & $2.198 \pm 0.011$
  & $0.033^{+0.007}_{-0.008}$
  &
  &
  &
\enddata

\tablenotetext{\,}{
Notes.---For each region, the table includes a list of the right ascension
($\alpha_{R}$) and declination ($\delta_{R}$) coordinates of the center of
the radio region, the angular distance between the centers of the X-ray (see
Table~\ref{tab2}) and radio regions $\Delta\Psi$, the 843~MHz flux divided
by the total 843~MHz flux ($\zeta$), the best-fit differential electron
spectral index $\Gamma$, curvature parameter $a$, and cutoff frequency
$\nu_{m}$, the value of $\chi^{2}$ per degree of freedom ($\chi^{2} / {\rm
dof}$), and the probability ($P_{\Delta \chi^{2}}$) that the difference
between the value of $\chi^{2}$ in Table~\ref{tab5} and the value in this
table is due to chance. The uncertainties, which include only the
statistical contributions, are reported at the 90\% confidence level.  These
uncertainties were used to compute the weighted mean values of $\Gamma$ and
$a$. Units of right ascension are hours, minutes, and seconds, and units of
declination are degrees, arcminutes, and arcseconds.
}

\tablenotetext{a}{
As described in the text, the spectral indices are not independent from
region to region.
}

\tablenotetext{b}{
Including curvature as a free parameter improves the quality of the fit for
every region.
}

\end{deluxetable}

%% file: tab5.tex
\begin{deluxetable}{cccccccccc}
\tablecaption{Best-Fit Parameters Using the Uncurved Model and Peak Radio
  Fluxes\label{tab5}}
\tablewidth{0pt}
\tablehead{
\
  & $\alpha_{R}$
  & $\delta_{R}$
  & $\Delta\Psi$
  & $\zeta$
  & 
  & 
  & $\nu_{m}$
  & 
  \\
Region
  & (J2000)
  & (J2000)
  & (arcmin)
  & ($\times 10^{-4}$)
  & $\Gamma$
  & $a$\tablenotemark{a}
  & ($10^{16}$~Hz)
  & $\chi^{2}/{\rm dof}$
}
\startdata
1
  & 15 04 07.1
  & $-$42 00 05
  & 0.68
  & 17
  & $2.13^{+0.08}_{-0.08}$
  & 0
  & $4.3^{+3.5}_{-1.7}$
  & 9.4/19
  \\
2
  & 15 04 07.7
  & $-$41 57 44
  & 1.04
  & 22
  & $2.11^{+0.07}_{-0.08}$
  & 0
  & $4.3^{+3.1}_{-1.6}$
  & 15.2/23
  \\
3
  & 15 04 04.4
  & $-$41 56 25
  & 1.73
  & 30
  & $2.10^{+0.07}_{-0.06}$
  & 0
  & $4.3^{+2.6}_{-1.5}$
  & 13.1/28
  \\
4
  & 15 04 04.6
  & $-$41 55 00
  & 0.95
  & 30
  & $2.05^{+0.06}_{-0.05}$
  & 0
  & $5.5^{+2.9}_{-1.7}$
  & 55.6/47
  \\
5
  & 15 04 06.1
  & $-$41 53 48
  & 0.67
  & 22
  & $2.01^{+0.05}_{-0.04}$
  & 0
  & $7.3^{+3.1}_{-2.0}$
  & 79.9/64
  \\
6
  & 15 04 02.6
  & $-$41 51 40
  & 0.77
  & 31
  & $2.05^{+0.04}_{-0.04}$
  & 0
  & $8.9^{+3.4}_{-2.3}$
  & 96.1/85
  \\
7
  & 15 03 59.9
  & $-$41 50 02
  & 1.03
  & 26
  & $2.12^{+0.05}_{-0.04}$
  & 0
  & $12.6^{+7.7}_{-4.3}$
  & 70.0/61
  \\
8
  & 15 03 58.6
  & $-$41 49 37
  & 0.53
  & 20
  & $2.04^{+0.05}_{-0.04}$
  & 0
  & $9.6^{+4.7}_{-2.9}$
  & 74.6/72
  \\
9
  & 15 03 54.1
  & $-$41 48 31
  & 0.72
  & 22
  & $2.04^{+0.04}_{-0.04}$
  & 0
  & $9.2^{+4.2}_{-2.6}$
  & 78.2/73
  \\
10
  & 15 03 50.5
  & $-$41 47 18
  & 0.52
  & 19
  & $2.04^{+0.04}_{-0.04}$
  & 0
  & $10.1^{+5.0}_{-3.0}$
  & 76.9/70
  \\
11
  & 15 03 47.8
  & $-$41 46 56
  & 2.35
  & 16
  & $2.12^{+0.04}_{-0.05}$
  & 0
  & $20.0^{+16.4}_{-8.0}$
  & 66.2/62
  \\
12
  & 15 03 41.7
  & $-$41 45 48
  & 3.02
  & 15 
  & $2.13^{+0.06}_{-0.06}$
  & 0
  & $15.0^{+14.1}_{-6.4}$
  & 32.9/38
  \\
13
  & 15 03 25.2
  & $-$41 43 45
  & 1.88
  & 17
  & $2.04^{+0.05}_{-0.06}$
  & 0
  & $7.5^{+4.6}_{-2.5}$
  & 43.1/43
  \\
Mean
  &
  &
  &
  &
  & $2.073^{+0.021}_{-0.020}$
  & 0
  &
  &
\enddata

\tablenotetext{\,}{Note.---See the caption for Table~\ref{tab4}. Units of
right ascension are hours, minutes, and seconds, and units of declination
are degrees, arcminutes, and arcseconds.}

\tablenotetext{a}{The curvature parameter was fixed at zero.}

\end{deluxetable}